\documentclass[12pt]{scrartcl}
\usepackage[utf8]{inputenc}
\usepackage{graphicx, array, amsmath, amssymb, lscape, multirow,listings, mathrsfs,verbatim,color,soul,xfrac}
\usepackage{parskip}
\usepackage{geometry}
\usepackage{lscape}
\usepackage{pdflscape}
\usepackage{setspace}

\usepackage{hyperref}
\hypersetup{
    colorlinks,
    citecolor=black,
    linktoc=all,
    filecolor=blue,
    linkcolor=black,
    urlcolor=black
}

\usepackage[round]{natbib}

\usepackage{tikz}
\usetikzlibrary{shapes}
\usetikzlibrary{fit}
\usetikzlibrary{chains}
\usetikzlibrary{arrows}

\usetikzlibrary{positioning,arrows.meta}
\tikzset{
     block/.style={rectangle, draw, fill=red!40, text width=6em,
                   text centered, rounded corners, minimum height=3em},
     arrow/.style={-{Stealth[]}}
     }
\tikzstyle{data} = [rectangle,fill=cyan!40!white,inner sep=3pt,minimum size=20pt,font=\fontsize{10}{10}\selectfont, node distance=1,text width=1.4cm,align=center]
\tikzstyle{eq} = [rectangle,fill=white,inner sep=3pt,minimum size=20pt,font=\fontsize{12}{12}\selectfont, node distance=1,align=center]
\tikzstyle{var} = [rectangle,fill=white,inner sep=3pt,minimum size=20pt,font=\fontsize{10}{10}\selectfont, node distance=1,text width=1.4cm,align=center,draw=black]
\tikzstyle{est} = [rounded rectangle,line width=0.2mm,draw=black,align=center]
\tikzstyle{module} = [circle,text width=1.4cm,line width=0.1pt,draw=gray,align=center]
\tikzstyle{plate} = [draw, rectangle, rounded corners, fit=#1]
\tikzstyle{wrap} = [inner sep=0pt, fit=#1]
\tikzstyle{caption} = [font=\footnotesize, node distance=0] %
\tikzstyle{plate caption} = [caption, node distance=0, inner sep=0pt,
below left=5pt and 0pt of #1.south east] %
\tikzstyle{factor caption} = [caption] %
\tikzstyle{every label} += [caption] %

\tikzstyle{latent} = [circle,fill=white,draw=black,inner sep=1pt,
minimum size=20pt, font=\fontsize{10}{10}\selectfont, node distance=1]
\tikzstyle{obs} = [latent,fill=gray!25]
\tikzstyle{margin} = [latent,dashed,draw=gray!40!black]
\tikzstyle{const} = [rectangle, inner sep=0pt, node distance=1]
\tikzstyle{factor} = [rectangle, fill=black,minimum size=5pt, inner
sep=0pt, node distance=0.4]
\tikzstyle{det} = [latent, diamond]
\newcommand{\plate}[4][]{ %
  \node[wrap=#3,#1] (#2-wrap) {}; 
  \node[plate caption=#2-wrap] (#2-caption) {#4}; 
  \node[plate=(#2-wrap)(#2-caption)] (#2) {}; 
}

\usepackage{xspace}
\newcommand{\fuday}{25\xspace}

\newcommand{\targetweight}{24\xspace}
\newcommand{\ppday}{32\xspace}

\newcommand*{\changes}[1]{#1} 

\title{Quantifying efficiency gains of innovative designs of two-arm vaccine trials for COVID-19 using an epidemic simulation model}

\author{Rob Johnson, Chris Jackson, Anne Presanis, \\ Sofia S. Villar, Daniela De Angelis}

\begin{document}

\maketitle

\begin{abstract} 
Clinical trials of a vaccine during an epidemic face particular challenges, such as the pressure to identify an effective vaccine quickly to control the epidemic, and the effect that time-space-varying infection incidence has on the power of a trial. We illustrate how the operating characteristics of different trial design elements  may be evaluated using a network epidemic and trial simulation model, based on COVID-19 and individually randomised two-arm trials with a binary outcome. We show that ``ring'' recruitment strategies, prioritising participants at imminent risk of infection, can result in substantial improvement in terms of power \changes{in the model we present}. In addition, we introduce a novel method to make more efficient use of the data from the earliest cases of infection observed in the trial, whose infection may have been too early to be vaccine-preventable. Finally, we compare several methods of response-adaptive randomisation, discussing their advantages and disadvantages in \changes{the context of our model} 
and identifying particular adaptation strategies that preserve power and estimation properties, while slightly reducing the number of infections, given an effective vaccine.
\end{abstract}

\clearpage

\section{Introduction}

Vaccine trials are \changes{still} in progress for SARS-CoV-2 and many vaccines are in development. So far, the trial designs proposed have been two-arm, individually randomised placebo-controlled trials, with the exception of the \cite{WHOR&DBlueprint2020}, which allows for new arms to be added. Vaccine trials in general have multiple, often competing, objectives which include establishing evidence on the efficacy of the vaccine, and conferring a health benefit to the trial participants as well as to the wider population \citep{Bellan2017}. Important decisions involved in designing any vaccine trial include the choice of trial population, whether randomisation takes place at an individual or cluster level, the comparator, and the primary endpoint definition. When a vaccine trial is conducted in the midst of an epidemic, these decisions must address the specific challenges of: identifying an effective vaccine as quickly as possible to control the epidemic \citep{Kahn2018a}; and the effect that variable infection incidence over space and time has on the power of a trial \citep{Camacho2015}. Efficiency gains that can address these challenges are a crucial topic of discussion, where ``efficiency'' includes: increasing power for a set number of participants or infections; reducing the required number of participants or infections for a given power; or reducing the time until conclusion for a given power. \citet{Kahn2018a}, \citet{Nason2016} and \citet{Kahn2020} discussed such efficiency gains in the context of the three diverse trials designed for the 2014--2016 epidemic of Ebola virus disease (EVD) \citep{Access2015,Kennedy2016,Widdowson2016}. 

In this paper, we \changes{focus on} three elements of trial design for testing vaccines in an epidemic that might improve power and efficiency: recruiting participants at highest risk of infection; making more efficient use of the data on participants who are infected earliest in the trial; and response-adaptive randomisation.

Recruitment of participants at highest risk of infection has been suggested \citet{Kahn2018a,Nason2016,Kahn2020} to be more efficient than random recruitment. If high-risk individuals can be confidently identified, recruiting from them can increase the number of cases that are observable in a fixed window of time. Power is directly increased by increasing the total number of events observed, but also indirectly by reducing the risks of the incidence rate considerably changing during the trial. ``Ring'' recruitment, where contacts of known cases are recruited, was implemented in a vaccine trial for EVD  \citep{Henao-Restrepo2017}. In the ongoing COVID-19 pandemic, many nations developed contact tracing systems to contain the spread of disease \citep{ECDC20202}, and contact tracing has been explored as a means for rolling out a tested vaccine \citep{MacIntyre2020.12.15.20248278}, but, to the best of our knowledge, contact tracing systems have not been formally used to define a recruitment strategy for vaccine trials. {Such a strategy to recruit patients for an ongoing treatment trial has been adopted to some success \citep{Cake2021}.} 

An important disadvantage of recruiting people at risk of imminent infection is that by the time patients are randomised, infection may already have occurred and thus cannot be prevented by the trial vaccine. Furthermore, even if infection has not occurred at randomisation, it might still occur before the vaccine induces a protective effect.   Including such individuals in a conventional analysis of the trial data would lead to underestimation of the vaccine efficacy. Alternatively,  participants observed to be infected before a particular time \changes{would commonly} be excluded (as in e.g. \citet{Henao-Restrepo2017}). However, since the time from (unobserved) infection to symptoms is variable, later cut-off times may exclude people who should have been included, thus reducing power, while earlier cut-off times will lead to underestimation of efficacy. Therefore, a more efficient way of using this information is desirable when \changes{ring recruitment} is used. 

Some adaptive designs \changes{may} improve efficiency, confer health benefits to trial participants, or in some cases achieve both \citep{Burnett2020, Pallmann2018}, \changes{including} in the context of COVID-19 treatment trials \citep{Stallard} and of vaccine trials \citep{Kahn2020}. Response-adaptive randomisation (RAR), in which the proportion of people randomised to a particular trial arm is modified based on accumulated data observed at an interim analysis, has been suggested \citep{Kahn2020,Scott2020} to have the potential to balance the competing objectives of health benefits to the trial participants, power and time until a conclusion is reached. \citet{Brueckner2018} compared different RAR designs for trials of treatments during an epidemic. However, the \changes{performance} of RAR has not been specifically quantified in the context of vaccine trials and its use in this context still remains debated \citep{Proschan2020, Villar2020}.

We present an epidemic simulation study to assess the impact on operating characteristics of the three proposed design elements in a specific plausible COVID-19-like situation.  This study also serves as a methodological \changes{example} of how epidemic simulation can be used to evaluate trial designs in any emerging epidemic, \changes{subject to the development of models} for the specific pathogen, epidemic and social context. 

First, we assess ring-type designs that recruit contacts of infected people in the context of individual randomisation. Second, we develop and evaluate a novel method to \changes{avoid excluding all data from early infections}, using weights that are estimates of the probability that a person was infected after vaccine-induced antibody response.
Finally, we compare various response-adaptive randomisation \changes{procedures}, under ring recruitment and downweighting early cases. We evaluate \changes{several operating characteristics} of four different frequentist and Bayesian methods for updating allocation probabilities.

In Section \ref{sec:simstudy}, we describe the network epidemic transmission model, the common characteristics of the trials we simulate, and the details of the design choices that we compare. In Section \ref{adapt3}, we present the estimated operating characteristics of each design. In Section \ref{sec:discuss} we conclude with a discussion of our findings, their limitations, and the potential for further work. Full details of the simulation model and trial mechanics are described in Appendix \ref{modelC19}, further details of the analysis method in Appendix \ref{exclusion_app}, supplementary results in Appendix \ref{suppres}, and code is available at \url{github.com/robj411/ADAGIO/COVID19}.

\section{Simulation study} \label{sec:simstudy}

\subsection{Network epidemic model}\label{sec:network}

We use a network model to simulate an epidemic occurring in a population in which the vaccine trial operates. There are two components to the model, following \citet{Hitchings2018}: the network that describes relationships between individuals, and the transmission model that describes the dynamics of disease via these relationships. Both the disease transmission network and the network for tracing contacts of identified cases are sub-networks of the relationship network.

\paragraph{Relationship network}
Our network is an undirected graph with vertices, or nodes, representing the $N_I$ individuals in the population of interest, and edges representing connections, or relationships, between individuals.  
These include relationships between people who know each other, defining the contact networks of the individuals (``known contacts''), and random relationships, between people who encounter each other only transiently (``transient contacts''). Transient contacts might include, for example, encounters between people who are travelling or in supermarkets, and are not defined to be part of contact networks. 

We consider three types of relationships: within household; in workplaces; and transient. Together, the household and the workplace edges are the \emph{known} edges that make up the known contacts. We assign a ``relationship weight'' of 0.1 to transient connections, compared to relationship weight 1 for non-transient relationships, to reflect the smaller probability of contact sufficient to enable transmission between transient connections than between acquaintances. 

Each individual has a set of attributes such as household and age ($<19$, 19-65, $>65$). Every individual in a household is connected to every other individual in the household. The age and household size distribution is taken from the United Kingdom 2011 Census \citep{Statistics2011}. People aged 19 to 65, and one fifth of people aged 65+, are connected to 15 other people, on average, via a workplace. The number of people in the workplace is reflective of the likely number of people with whom an infrastructure is shared, rather than the number of colleagues. Individuals within a workplace are completely connected to one another.

Finally, random edges are added, to allow around ten transient connections per person, i.e. potential transmission encounters that would not be recalled or anticipated through contact tracing. The result is an average of 20 connections per person, of which ten have a weight of 1 and ten have a weight of 0.1. An example is shown in Figure \ref{contactnetworkc19}. Full details of the network are given in Appendix \ref{modelC19}.
\begin{figure}
\begin{center}
\includegraphics[width=0.5\textwidth]{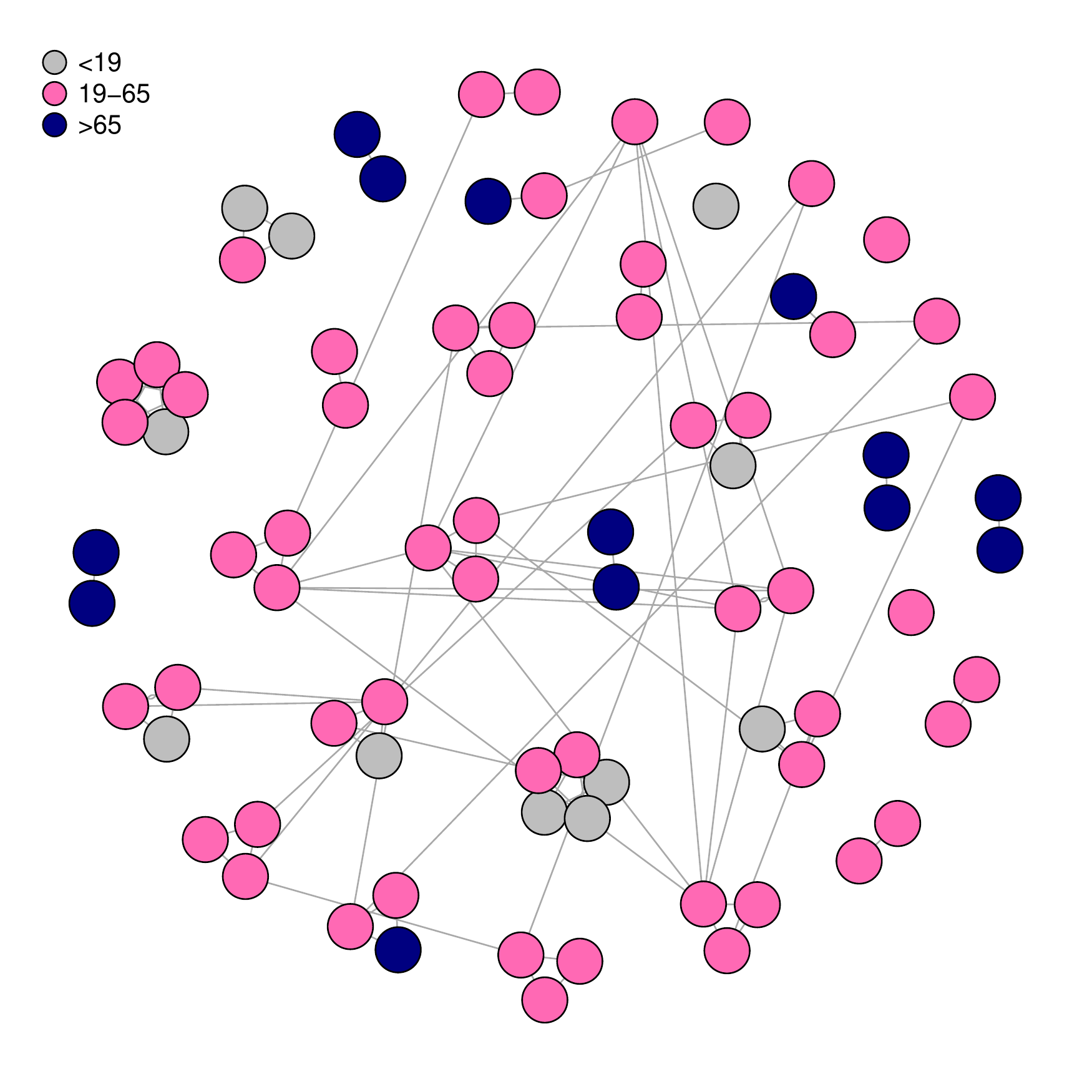}
\end{center}
\caption{\small Example relationship network showing 80 people. ``Known'' edges between housemates and colleagues are shown. Individuals are coloured by age group and clustered into households.}
\label{contactnetworkc19}
\end{figure}

\paragraph{Disease and trial state transitions}

\begin{figure}[h]
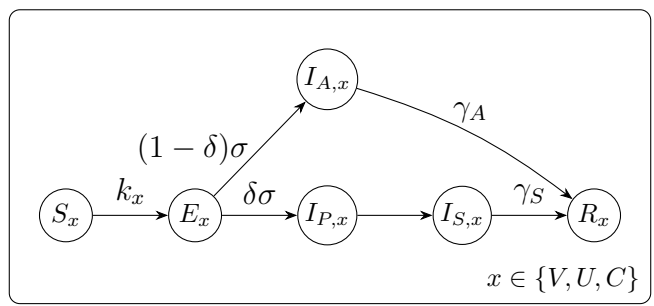

  \centering
  \tikz{ %
    \node[latent] (S) {$S_{x}$} ; %
    \node[latent, right=of S] (E1) {$E_{x}$} ; %
    \node[latent, right=of E1] (L1) {$I_{P,x}$} ; %
    \node[latent, above=of L1] (L2) {$I_{A,x}$} ; %
    \node[latent, right=of L1] (I1) {$I_{S,x}$} ; %
    \node[latent, right=of I1] (R) {$R_{x}$} ; %
    \draw [arrow] (S) -- node [above] {$k_x$} (E1);
    \draw [arrow] (E1) -- node [above] {$\delta\sigma$} (L1);
    \draw [arrow] (E1) -- node [left] {$(1-\delta)\sigma$} (L2);
    \draw [arrow] (L1) -- node [above] {} (I1);
    \draw [arrow] (I1) -- node [above] {$\gamma_S$} (R);
    \draw [arrow] (L2) edge[bend left=10] 
    node [above] (T6) {$\gamma_A$} (R);
   \plate[inner sep=0.25cm, yshift=0.12cm] {plate3} {(S) (L2) (R)} {$x\in\{V,U,C\}$}; %
  }
  \caption{\small Disease-state transition model for members of the population who are not enrolled ($U$) and those enrolled and vaccinated ($V$) and those enrolled to the control group ($C$). Arrows show possible transitions between states, labelled by the rates.
  }
\label{communities}
\end{figure}

Individuals' disease states and possible transitions are described by a compartmental SEIR model (Figure \ref{communities}), with a structure similar to \citet{Camacho2015} and \citet{Danon2020}. The possible states are $S$ Susceptible, $E$ Exposed (infected but not infectious), $I_A$ Infectious and asymptomatic, $I_P$ Infectious and pre-symptomatic, $I_S$ Infectious and symptomatic and $R$ Removed. Trial enrollment states are represented by the subscript $x$, where $x = U$ is not enrolled, $x = V$ is enrolled and vaccinated, $x = C$ is enrolled and in the control arm.

Every individual $i$ starts susceptible and unenrolled, in state $S_U$, except the index case who starts in $E_U$. The infection hazard $k_{x}(i)$ for an individual $i$ in susceptible state $S_x$ is a function of the per-contact transmission rate (see Appendix \ref{modelC19}), their contact network, their vaccination status $x_i$ and the vaccine efficacy (VE) $0 \leq \eta \leq 1$, defined as the percent reduction in attack rate for vaccinated people compared to unvaccinated people, assuming the vaccinated population have reached the maximum state of protection they are capable to reach  \citep{Weinberg2010,Shim2012}. We equate this state with vaccine-induced antibody response \citep{Hudgens2004}.  An infected individual in the exposed state $E_x$ becomes infectious but asymptomatic with probability $1 - \delta$, moving to state $I_A$ at rate $(1 - \delta)\sigma$. The remaining proportion $\delta$ become infectious and pre-symptomatic, moving to state $I_P$ at rate $\delta\sigma$. The transition rate $\sigma$ corresponds to an incubation period $\xi \sim 2 + \Gamma(\text{shape}=13.3,\text{rate}=4.16)$ \citep{Li2020} and we set $\delta=0.8$ \citep{Buitrago-Garcia2020}.

The asymptomatic individuals remain asymptomatic and therefore stay in $I_A$ for their whole infectious period, $\psi \sim 1+\Gamma(\text{shape}=1.43,\text{rate}=0.549)$ \citep{Li2020}, corresponding to a rate $\gamma_A$, before moving to the removed state $R$. The pre-symptomatic individuals in $I_P$ move to $I_S$, after a deterministic time of 1 day \citep{Kucharski2020}. Symptomatic individuals remain in $I_S$ for the remainder of their infectious period, $\psi - 1 \sim \Gamma(\text{shape}=1.43,\text{rate}=0.549)$, corresponding to rate $\gamma_S$, before moving to the removed state $R$.  \changes{For simplicity, we assume} that symptomatic persons do not leave their homes, so that an infectious person in $I_A$ or $I_P$ can infect their home contacts, their work contacts and their random contacts, but an infectious person in $I_S$ can only infect their home contacts. Transitions to state $R$ imply removal from the infectious population: this can be due to death, hospitalisation (and hence isolation), or recovery.

See Appendix \ref{modelC19} for full details of the transmission model.

We simulate $500$ households, corresponding to populations of around 1,000 individuals. Our simulated trials operate on a time unit of one day, and have total duration of the order of 100 days. We assume that one contact network (corresponding to one index case) is initiated every day. Initiation is the moment where all nodes in the network are in state $S_U$ except one individual who is on their first day in state $E_U$. Enrolment can only begin when this individual reaches state $I_S$. Each trial participant who is enrolled has as their reference day the day on which they were enrolled. We simulate a vaccine efficacy of either no effect $\eta = 0$ or a positive effect $\eta = 0.7$. 

\subsection{Trial design}

The elements of trial design we explore are specific to an infectious disease with the dynamics and means of spread of EVD or COVID-19 (i.e. through close person-to-person contact) \changes{with} the primary endpoint \changes{defined as} a reverse transcriptase polymerase chain reaction (RT-PCR)-confirmed diagnosis of current infection. In addition, our trial is for a single-dose vaccine whose time to development of antibodies is fast, i.e., a (hypothetical) disease-specific antibody test would go from negative to positive within 15 days, for all participants with high confidence \citep{Marzi2015,Poland2020}. We use antibody response as a proxy for a vaccine-induced protective immune response in our simulations. 

{We develop a trial design step by step. First, we design a ``base case''. Then, we choose one element of the design to change (e.g. the recruitment strategy from random to ``ring'') and compare the new design to the original. We then take forward the better-performing design and make a comparison on another element of the design.}
In the starting ``base case'' for our simulations, participants are recruited at random; the allocation probabilities are fixed and equal (fixed randomisation, FR); the final follow-up time is \fuday~days after randomisation, similar to the follow-up time of 21 days used for EVD \citep{Hitchings2017,Access2015}; the trial terminates once \targetweight confirmed cases have been observed {(determined to achieve 80\% power for a vaccine efficacy of 0.7)}, and is analysed assuming everyone receives \changes{the intervention they were randomised to}.

In the base case, similar to \citet{Henao-Restrepo2017}, who excluded those exhibiting symptoms within 10 days in their trial for an EVD vaccine, and \citet{WHOR&DBlueprint2020}, who propose 14 days for COVID-19, we exclude cases that display symptoms within nine days of randomisation. Our rationale for 9 days aligns with our assumed time to vaccine-induced antibody response and {the estimated} incubation period {for SARS-CoV-2 \citep{Li2020}}. \changes{Notice that} the Blueprint uses a time-to-event outcome, whereas we use a binary outcome (PCR-confirmed diagnosis of COVID-19), \changes{which enables us to use well-known methods} of response-adaptive randomisation, and for which we have a maximum follow-up time of \fuday~days.

The primary endpoint in our simulations corresponds to a disease endpoint, where any person who becomes symptomatic 
is PCR tested and diagnosed.  We assume that through surveillance and self-reporting, symptomatic participants correspond to confirmed cases, i.e.  all symptomatic people report their symptoms and onset date accurately, all tests have perfect accuracy, and each symptomatic person's positive result is available by the time of the end of their follow-up period of \fuday days. In reality, some people test positive for SARS-CoV-2 but show no symptoms. In our simulations,  these people are infectious but not symptomatic. In terms of the network epidemic model, they continue to behave as if they are not infectious, so they retain contacts whom they can infect over their infectious period. In terms of the trial, they are not diseased and so are counted in the group of trial successes. 
An ``infection endpoint'' \citep{Hudgens2004}, where only an ``uninfected'' outcome is a success of the vaccine, would \changes{require} routinely testing all participants.

We use an event-driven approach \citep{Schoenfeld1983} for trial size determination. The trial is terminated following the accumulation of a prespecified number of confirmed cases, rather than a prespecified number of enrolled participants (which corresponds to a fixed sample size).
In the comparisons we make, all designs employ randomisation at the individual level and compare to a placebo control, as suggested by \citet{Kahn2020}.

\subsection{Recruitment}

We compare random recruitment to the ring-recruitment strategy employed in \citet{Henao-Restrepo2017}. Participants are eligible for enrolment when someone in their contact network is confirmed as a ``case''. We use ``ring'' with reference to the method for recruitment, and we use individual-level randomisation, whereas other implementations of the strategy have used cluster randomisation \citep{Henao-Restrepo2015}. \changes{In cluster randomisation an individual is eligible only if the contact network of its index case is eligible, based on the trial's inclusion criteria. Therefore individual randomisation can enroll more participants per index case as eligibility is individual.} We define the ring as consisting of contacts and contacts of contacts. These contacts are found through contact tracing, as described by the \citet{ECDC20202}. 
In Appendix \ref{recruitment_appendix}, we show how the success of the ring recruitment method depends on the ability of contact tracing to identify those at imminent risk of infection.

\subsection{Weighted exclusion}\label{weighting_app}

Explicit mention of time to vaccine-induced antibody response is not often made in the definition of vaccine efficacy, although exclusion rules often make reference to this time: participants who are confirmed as cases within a certain number of days of randomisation are excluded from the analysis (``binary exclusion''), as they are assumed to have been infected either before randomisation or before the vaccine has a chance to take effect \citep{Dean2019,Henao-Restrepo2017}. 

However, by excluding observations, we lose some information. Our ideal endpoint is whether or not a participant became infected after randomisation and vaccine-induced immune response, where participants who became infected before are excluded from the trial. We typically do not know the day on which a vaccinated person develops infection- or vaccine-induced antibodies. However, we can observe whether or not a person is symptomatic, at which point infection can be confirmed through laboratory testing. Therefore, for the analysis at the end of the trial, we propose supplementing the primary endpoint -- the observed infection status at day~\fuday{}  -- with a retrospective exclusion criterion: we use our knowledge of the disease (e.g. the incubation period) and the day a person becomes symptomatic to weight their inclusion in the analysis. 

We define our primary endpoint for the final analysis as \changes{PCR-confirmed, symptomatic infection} 
and, in addition, the day of symptom onset relative to the day of randomisation, which informs a retrospective exclusion criterion. This exclusion of participants is expressed as an ``inclusion weight'' between 0 and 1, computed at each round of analysis, and applied retrospectively, as if we had excluded a participant from the beginning (``continuous inclusion''). When confirmed cases are weighted, the number of cases becomes the \emph{effective number of cases}, and the sample size the \emph{effective sample size}.

The probability that a person whose symptoms began after vaccination was infected after vaccine-induced antibody response depends on the effect of the vaccine. If the vaccine is effective, then this probability is smaller than it would be if the vaccine had no effect. Therefore, as an additional development, we estimate the vaccine efficacy and the inclusion weights together iteratively, so that both would be updated if (as in a response-adaptive design) we were to recalculate them as results accumulate. See Figure \ref{veweight} in Appendix \ref{exclusion_app} for an illustration of how the weights are obtained, and for the mathematical derivation. We assess this weighting method in terms of power and type 1 error compared to binary exclusion.

\subsection{Response-adaptive randomisation}\label{adaptations}

Let $\pi_v$ be the probability of being allocated to arm $v$, where $v$ is 0 for control and 1 for experimental. Thus far we have considered fixed-randomisation (FR) designs, in which the allocation probabilities are fixed and equal ($\pi_0=\pi_1=0.5$) throughout the trial. In a response-adaptive randomisation allocation probabilities are updated at pre-specified moments in the trial using accumulated data according to a pre-specified rule.

We set the frequency of adaptation to be every \fuday days: all data accrued up to the adaptation day (up to the maximum follow-up time of \fuday days post randomisation) are used to generate the probabilities. Therefore, the allocation probabilities are updated after groups of participants of random size, rather than individuals, and the first group acts as a ``burn-in phase'' where there is an equal probability of receiving each arm.  We present frequentist and Bayesian methods to generate $\pi_1$. The methods all require first estimating $p_0$ and $p_1$: the probability of being uninfected up to the maximum follow-up time for the control and experimental arms, respectively. We denote the estimates $\hat{p}_0$ and $\hat{p}_1$.

The probabilities are estimated as the number of successes over the total number of observations; $\hat{p}_v=1-f_v/N_v$, where $f_v$ is the effective number of confirmed cases for arm $v$, and $N_v$ its effective sample size \citep{Hu20062}.

\subsubsection{Frequentist response-adaptive}\label{freq}

In the Rosenberger et al. method \citep{Rosenberger2001}, favourable outcomes are optimised subject to a power constraint, so that $\rho_{Ros}=\sqrt{\frac{\hat{p}_1}{\hat{p}_0}}$ is the optimal randomisation ratio of experimental to control. Then the allocation probability to the experimental arm is:
$$\pi_{1}=\frac{\rho_{Ros}}{\rho_{Ros}+1}=\frac{\sqrt{\hat{p}_1}}{\sqrt{\hat{p}_1}+\sqrt{\hat{p}_0}}.$$

We also consider the Neyman method, which is designed to maximise power \citep{Rosenberger2001}, by setting 
$$\pi_1= \frac{\sqrt{\hat{p}_1(1-\hat{p}_1)}}{\sqrt{\hat{p}_0(1-\hat{p}_0)}+ \sqrt{\hat{p}_1(1-\hat{p}_1)}} .$$

For example, if the infection rate for the control arm is 80\% ($\hat{p}_0=0.2$) and that in the experimental arm is 20\% ($\hat{p}_1=0.8$), then the Rosenberger et al. method would allocate participants in a 2:1 ratio favouring the experimental arm ($\rho_{Ros} = \sqrt{4} = 2$, and $\pi_1= 2/3$). The Neyman rule has $\pi_1= 0.5$ (and $\rho_{Ney}=1$). For these methods, in the case that $\hat{p}_0(1-\hat{p}_0)=0$ or $\hat{p}_1(1-\hat{p}_1)=0$, we set $\pi_0=\pi_1=0.5$.

\subsubsection{Bayesian response-adaptive (Thompson sampling)}\label{bayes}

The Bayesian methods define the allocation probability in terms of the posterior distributions of $p_i$ given a uniform prior and the observed data, $\text{Beta}(1+N_v-f_v,1+f_v)$. Then $\pi_i$ is estimated by sampling as \citep{Thall2007}
\begin{equation}\label{tseq}\pi_1=\frac{\text{Pr}(p_1>p_0)^\phi}{\text{Pr}(p_1>p_0)^\phi+\text{Pr}(p_1<p_0)^\phi},\end{equation} where we define a tuning parameter $\phi$, which tempers the speed with which the allocation probability can reach extreme values (0 or 1). For Thompson sampling (TS), we set $\phi=1$, so $\pi_1$ is just $\text{Pr}(p_1>p_0)$, and for TS with tuning (TST), $\phi=j/e$, with $j$ the day of the current update, and $e$ the trial's expected total duration. One might instead choose to adapt according to number of cases seen, so that $e=$ \targetweight effective cases. $\phi$ therefore takes the value 0 at the beginning of the trial and goes to 1 as the trial progresses. 

The Thompson sampling algorithm has a possibility of generating extreme allocation probabilities. While tuning limits this, we find that the TST method still tends to 1 over few adaptations. We therefore \changes{set limits to the allocation probability: we use a value of 0.8 if Equation \eqref{tseq} returns an allocation probability above 0.8, and we use a value of 0.2 if Equation \eqref{tseq} returns an allocation probability below} 0.2 for both implementations of Bayesian response-adaptive randomisation. We additionally terminate the trial early and conclude efficacy if Equation \eqref{tseq} returns a value of 0.99 \citep{Brueckner2018}.

\paragraph{Time trends}

The epidemic unfolding in real time can give rise to temporal trends in incidence of the disease among participants, \changes{also referred to as} ``patient drift'' \citep{Proschan2020,Villar2020}. Patient drift affects all arms in the same way, and might be induced by a natural increase or decrease in incidence, or a step change due to government policy on social contact, or a change in the recruitment process. As both the adaptive trial design and the epidemic change over time, we must account for time dependencies of disease exposure when inferring the effect of the experimental vaccine. 

Here, we use randomisation-based inference 
as described by \citet{Simon2011}: we resample the data in order to generate a new null distribution for the test statistic to which to compare the one we compute. We present the resulting resulting power and type 1 error rates alongside the uncorrected values \changes{from standard testing}.

\subsection{Evaluation}\label{evaluation}

We simulate $N_T$ trials, where each ``trial'' involves independent networks -- as many as are required to achieve a particular total effective number of cases. The null hypothesis is no effect of the vaccine, $H_0: \eta = 0$ and the alternative hypothesis is a positive vaccine efficacy, $H_1: \eta = \eta_1$ for a certain $\eta_1>0$.

We report operating characteristics including the number of people enrolled and the number of confirmed cases, the power, the estimated VE, and the type 1 error rate, alongside the details of the design. The duration of the trial is reported in days, and an average of $N_P$ participants are enrolled per day, according to the properties of our simulated network and enrolment rate. 

The VE, $0\leq \eta \leq 1$ is estimated as
$$\hat{\eta} = 1 - \left.\cfrac{f_1}{N_1}\right/\cfrac{f_0}{N_0},$$
where $f_v, v = 0, 1$ is the effective number of cases in arm $v$ and $N_v$ the effective number of participants in arm $v$. The VE is estimated using all simulations under the positive effect, whether or not the trial realisation concluded efficacy.

 Power is the probability of correctly rejecting the null hypothesis $H_0$, and is estimated as the proportion of simulations under the alternative $H_1$ for which the $H_0$ is rejected.  Type I error rate is the probability of incorrectly rejecting $H_0$ when it is true, and is estimated as the proportion of simulations under the null for which the null hypothesis was rejected (see Appendix \ref{exclusion_app} for details).  Other results presented (e.g. the numbers of people enrolled and confirmed as cases) are computed under the alternative unless stated otherwise.

Additionally, we report a novel metric to evaluate  the different vaccine trial designs: the  ``prevented exported infections''. It is defined as the reduction in expected number of infection events of people not in the index case's contact network, for 100 contact networks, comparing a trial realisation with no vaccine effect with one with a positive effect of $\eta_1$, in the case that the vaccine prevents infection as well as disease. While we do not expect this metric to be predictive of actual numbers of infections occurring, the relative numbers between methods are indicative of the trials' possible or probable effects on the wider epidemic.

\section{Results}\label{adapt3}

The results in this section are from simulations of $N_T = 10,000$ trials, where the alternative hypothesis positive effect is set to $\eta_1 = 0.7$. We assume one contact network is enrolled per day. Given the network sizes and enrolment rate we assume, an average of $N_P = 32$ people are enrolled per day.

\subsection{Recruitment}

Where recruitment is random, rather than through contact tracing, we have to recruit many more participants (Table \ref{covid}), and many more people in the general population need to become infected,\footnote{``Exported infections'' is the expected number of infections of people outside the index case's contact network, for 100 contact networks. ``Prevented exported infections'' is the difference between exported infections assuming a vaccine efficacy of 0.7 and a vaccine efficacy of 0 in the case that the vaccine prevents infection as well as disease.} in order for the requisite number of infections to be observed among those recruited. In addition, fewer exported infection events are prevented. 
We carry forwards the ring-recruitment design as the ``base case'' for further comparisons.

\newgeometry{,vmargin=2cm,hmargin=1cm}
\begin{landscape}
\begin{table}[ht]
\centering
\caption{\small Comparison of designs where participants are recruited following the ring strategy vs. recruited at random. The trial follows the FR design with a follow-up time of \fuday days. The trial ends when an effective number of \targetweight cases have been observed. Standard deviations for 10,000 simulations  in brackets.}
\label{covid}
\centering
\begin{tabular}{lp{1.95cm}p{1.7cm}rrp{1.7cm}p{1.7cm}p{1.7cm}}
  \hline
Recruitment  & Number of participants & Number of confirmed cases & Vaccinated & Power & Type 1 error & VE\newline estimate  &  Prevented exported infections\footnotemark[2] \\ 
  \hline
  Random & 11275 (2345) & 34 & 5637 & 0.74 & 0.04 & 0.63 (0.18) & 2.53 \\ 
  Ring & 1929 (583) & 45 & 965 & 0.73 & 0.04 & 0.58 (0.19) & 6.29 \\ 
 \hline
\end{tabular}
\end{table}

\begin{table}[ht]
\centering
\caption{\small Comparison of designs where the binary endpoint has a binary weight or a continuous weight. Participants are recruited following the ring strategy.  The trial follows the FR design with a follow-up time of \fuday days. The trial ends when an effective number of \targetweight cases have been observed for the continuous weight and 26 for the binary weight, in order to achieve comparable trial sizes in terms of the number of participants. Standard deviations for 10,000 simulations in brackets. 
}
\label{binary}
\begin{tabular}{lp{1.7cm}p{1.7cm}rrp{1.7cm}lp{1.7cm}p{1.7cm}}
  \hline
 Weighting & Number of participants & Number of confirmed cases & Vaccinated & Power & Type 1 error  & VE estimate & Number of participants (null) & Prevented exported infections\footnotemark[2] \\ 
  \hline
  Binary & 2083 (609) & 49 & 1041 & 0.75 & 0.04  & 0.58 (0.18) & 1302 (430) & 6.25 \\ 
  Continuous & 2136 (624) & 50 & 1068 & 0.82 & 0.05  & 0.64 (0.19) & 1277 (426) & 6.35 \\ 

   \hline
\end{tabular}
\end{table}
\end{landscape}
\restoregeometry

\newgeometry{,vmargin=2cm,hmargin=1cm}
\begin{landscape}

\begin{table}[ht]
\centering
\caption{Comparison of response-adaptive designs. The outcome has a continuous weighting. Participants are recruited following the ring strategy. The final follow-up time is \fuday days. The trial ends when \targetweight effective cases have been observed. Standard deviations {for 10,000 simulations}  in brackets. Correction for time trend uses the resampling method of \citet{Simon2011}. Bold type indicates the recommended method.}
\label{rar}
\begin{tabular}{lp{1.9cm}p{1.7cm}p{2.4cm}rrp{1.7cm}p{1.7cm}p{1.9cm}lp{2.3cm}}
  \hline
Adaptation  & Number of participants & Duration (days) & Number of confirmed cases & Vaccinated & Power & Power (corrected) & Type 1 error & Type 1 error (corrected) & VE estimate & Prevented exported infections\footnotemark[2] \\ 
  \hline
  Ney & 1947 (551) & 85 (17) & 54 & 816 & 0.83 & \textbf{0.76} & 0.06 & \textbf{0.04} & 0.67 (0.2) & 4.62 \\ 
  Ros & 2147 (630) & 92 (20) & 57 & 1083 & \textbf{0.82} & 0.79 & \textbf{0.05} & 0.04 & 0.64 (0.18) & 5.38 \\ 
  TST & 2032 (638) & 88 (20) & 51 & 1261 & 0.77 & \textbf{0.76} & 0.04 & \textbf{0.04} & 0.64 (0.19) & 6.62 \\ 
  TS & 1799 (740) & 81 (23) & 45 & 1148 & 0.80 & \textbf{0.74} & 0.04 & \textbf{0.05} & 0.67 (0.21) & 5.99 \\ 
  FR & 2137 (622) & 91 (19) & 57 & 1068 & \textbf{0.82} &  & \textbf{0.05} &  & 0.64 (0.19) & 5.50 \\ 

   \hline
\end{tabular}
\end{table}
\end{landscape}
\restoregeometry

\subsection{Weighted exclusion}

In Table \ref{binary}, we \changes{show results that suggest that} by downweighting (``Continuous'') inclusion, rather than applying a binary rule, there is an increase in power \changes{of 0.07}, and the VE estimate is closer to the true value of 0.7. The gain in power is in large part due to accounting for the vaccine efficacy when determining which early cases are likely to have been infected before randomisation.  Other operating characteristics are similar between the methods.

\subsection{Response-adaptive randomisation}\label{rarres}

The comparison between the fixed-randomisation (FR) trial design developed up to now and the suite of response-adaptive designs is shown in Table \ref{rar}. We fixed the number of effective cases observed in the trial population to a total weight of \targetweight so that powers were comparable, which sets the health cost to the trial participants for the Neyman, Rosenberger et al. and FR methods. 
We can then trade off the power against time to conclude.  The Neyman method, which by design maximises power, in fact has much lower power once we correct for patient drift using a randomisation based approach. The Rosenberger et al. method is most similar to the fixed and equal randomisation design, matching it in terms of number of participants, participant allocation, type I error and power, and number of participants vaccinated. 

The Thompson sampling methods benefit from stopping early when efficacy can be concluded; the TS design is expected to be shortest among all designs (see Table \ref{rarts} for operating characteristics when the trials do not terminate early). Thus the health cost to the trial participants for these methods is not prespecified and the number of cases among participants must also be taken into account when evaluating the methods. The Thompson sampling methods (TST and TS) allocate more participants to an effective vaccine than control when it exists. As a result, there are fewer infections exported from the network, and the power is lower (both with and without correction). See Figure \ref{time_trend_fig} for operating characteristics under different trends, and Figure \ref{allocation_probability} for trajectories of allocation probabilities.

Table \ref{rar} illustrates how adaptive designs can be compared and how one might choose a design given the current circumstances: that, at a cost of some power, a design can be chosen that will vaccinate more people, if the vaccine is effective. This might be preferable in circumstances where infection rates are high. On the other hand, where infection rates are declining, a trial that maximises power might be preferable, since it would be more challenging to observe cases quickly. Such a design likely would not prioritise vaccination, prioritising instead information gain in order to increase the chance of identifying an efficacious vaccine. In Appendix \ref{fixed_trial_duration_section} we compare the same designs assuming instead that the trial must conclude within a certain number of days. Alternatively, designs could be compared in terms of the number of cases, the number of vaccinations, and duration, where all designs achieve the same power.

\section{Discussion}\label{sec:discuss}

Using simulation from a network epidemic model for COVID-19 with an
embedded vaccine trial, we have \changes{illustrated} the potential
efficiency gains from three innovative two-arm trial design and
analysis elements.    These elements \changes{are designed to} address the requirement in an
epidemic to observe as many events as quickly as possible, both for
control of the epidemic and information gain in the context of \changes{a highly} variable
and potentially \changes{low} incidence. \changes{The utility of each of these elements will depend on the disease and context and should be assessed through simulation.}

\changes{The first element,} ring recruitment, which prioritises individuals at imminent
risk of infection, has been shown to substantially improve power and
efficiency.    Our \changes{proposed} weighted analysis method
makes more efficient use of the available data:  \changes{reducing} bias compared
to fully including the data from people infected a short time after
randomisation, who the vaccine may not have had a chance to protect,
while \changes{gaining} power compared to completely excluding these data. While
response-adaptive randomisation \changes{may not offer a notably superior balance in terms of competing goals} in the two-arm trials considered in this study, we
nevertheless found that, given a moderately effective and safe vaccine, the adaptation method of \cite{Rosenberger2001} was comparable to a fixed
randomised design, preserving type I error and power while vaccinating
slightly more people in the trial.     
Each of these three design elements could independently increase power, efficiency \changes{or patient benefit} of a vaccine trial in particular contexts.  Furthermore,
the combination of all three simultaneously has
the potential to improve a vaccine trial in an epidemic context both
from the information gain and the health benefit perspective. We
believe this conclusion is valuable given the limited scope for
efficiency and participant benefit improvements that two-arm trials
usually have.

Response-adaptive designs require an outcome that is
observable soon after randomisation, which can be achieved with a ring-recruitment strategy. Ring recruitment requires an efficient contact-tracing infrastructure to
enable recruitment of participants at imminent risk of infection.  Such
contact tracing might be embedded in a national surveillance system
aimed at containment, or might be part of the trial protocol.  
The ring design depends on the ability to anticipate among whom new infections will occur: specifically, if new
cases occur among known contacts of cases, which might be ascertained through comparison of contacts traced and case registries.  A COVID-19
treatment trial in the UK has successfully used the UK's National Health
Service contact tracing data to enhance their recruitment \citep{Cake2021}.

The success of the ring design depends also on the time taken to trace contacts relative to disease dynamics. Our simulations assumed that, on
average, it takes ten days to identify and enroll a whole contact
network, including the time for the index case to be
confirmed (Table \ref{parameters}).  The sooner participants are enrolled after their index case becomes infectious, the more chance there is for an efficacious vaccine to confer protection, as vaccination occurs earlier relative to the time that the participant is at risk. This timing will depend also on when infectiousness begins (which, for COVID-19, is before symptom presentation) and the disease's incubation period.
Fast enrolment relative to disease progression enhances information gain per participant as well as the potential health benefit to those in the experimental arm. If contacts cannot be traced fast enough, ring recruitment would not be an appropriate method. 
\changes{Some debate about the limitations of the ring design of \citet{Henao-Restrepo2017} has been expressed, including the fact that it was a cluster randomised trial, and so ``subject to the same biases as other cluster randomised trials''} \citep{RidMiller2016}. \changes{Here, instead, we have used individual randomisation.}

Most of the response-adaptive randomisation designs we considered
incurred a penalty in power. The penalty increased when we controlled for bias due to patient drift with re-randomisation \citep{Simon2011}.  The more the allocation deviates
from equality, the greater the design's intended benefit but the
larger penalty in terms of power. Bounding the allocation probabilities
between 0.2 and 0.8 guards against very severe penalties, and would make the design more acceptable to stakeholders. \changes{We recommend that the trade-offs
between strictly preserving type I error and the resulting power loss when using randomisation based tests are
considered carefully at design stage through extensive simulations.}
Alternative corrections, for example stratification
\citep{Chandereng2019}, might prove less costly. The two-arm
trials we considered give some insights into what a multi-arm response-adaptive design could offer.  In a two-arm trial, power given a fixed number of participants can only be increased at the expense
of participant benefit \citep{Williamson2017,Villar2015a}. In
a multi-arm trial, worse-performing experimental arms can be
deprioritised in favour of other arms
\citep{Tymofyeyev2007}.  In our simulations, type 1 error is not much
inflated for the Thompson sampling methods since, under the null (i.e.
no vaccine effect), \targetweight effective cases are typically observed soon after the end of the equal-randomisation burn-in phase (Figure
\ref{allocation_probability}), so that the allocations are not
very imbalanced. For a Thompson sampling design that adapts earlier
relative to its end time, we would expect to see an inflated type 1 error, as in Figure \ref{time_trend_fig}.

We expect that the two-layer ring designs presented here would not suit
a trial for a two-dose vaccine for COVID-19, since people at imminent
risk of infection are recruited, and so the majority of infections would
occur before they could be prevented by a second dose.  Thus any
estimates of efficacy would only describe the efficacy of the first
dose.  Adaptive designs may also be less appealing for a two-dose
vaccine, since the disease outcome after two doses would take longer to
observe.

In our simulations we consistently underestimate the vaccine effect, \changes{and this is more pronounced for the ring recruitment designs}.  
Our method of downweighting rather than excluding the earliest cases
(which may not have been vaccine-preventable) is designed to improve
power compared to excluding these cases, and controls bias compared to
including them all.  Any remaining bias could be
controlled further by decreasing the weight assigned to early cases, at
the cost of reducing power.

\changes{We have illustrated how simulation might be used to compare different designs and analysis options, in an approach similar to} \citet{Hitchings2018}. In practice, the network epidemic model must be
specific to the particular setting, taking into account contact
structures and governmental policies, as both network and epidemic dynamics will impact the trial designs' operating characteristics. Through simulation, the design rules, such as the follow-up time and the requisite
effective number of cases to achieve the desired power, can be established. Additionally, sensitivity to the structural and parametric assumptions underlying the network, epidemic, and trial models can be evaluated. \changes{To fully capture that the trial occurs within a real-life epidemic,} 
the individual simulated contact networks could be embedded in a single, connected network on which the epidemic spreads, \changes{rather than simulated as independent units}.
Embedding the trial
simulation more comprehensively in an epidemic model, where the trial participants from different contact networks interact with each other and where the trial
can impact on the epidemic, would permit a formal quantification of
the benefits and limitations of different design choices in different epidemic settings,
as in \citet{Bellan2017}. Such an analysis would enable a realistic
assessment of the impact of more complex time trends at different stages
of an epidemic, and of the potential impact on the epidemic of designs
that vaccinate more people.

\enlargethispage{\baselineskip}

\section*{Acknowledgements}

This research is funded by the Department of Health and Social Care using UK Aid funding and is managed by the NIHR (grant number PR-OD-1017-20006). The views expressed in this publication are those of the author(s) and not necessarily those of the Department of Health and Social Care. CJ, AP and DDA were also funded by the UK Medical Research Council programme MRC\_MC\_UU\_00002/11 and SV by the UK Medical Research Council programme MRC\_MC\_UU\_00002/15. The authors also thank David Robertson, Thomas Jaki, Marc Lipsitch, Rebecca Kahn, Ben Cooper, Kendra Wu and Peter Horby for helpful discussions.

\clearpage
\bibliography{adagio,adagio2}

\begin{thebibliography}{}

\bibitem[Bellan et~al., 2017]{Bellan2017}
Bellan, S.~E., Pulliam, J. R.~C., {Van Der Graaf}, R., Fox, S.~J., Dushoff, J.,
  and Meyers, L.~A. (2017).
\newblock {Quantifying ethical tradeoffs for vaccine efficacy trials during
  severe epidemics}.
\newblock {\em bioRxiv}.

\bibitem[Box, 1954]{Box1954}
Box, G. E.~P. (1954).
\newblock {Some theorems on quadratic forms applied in the study of analysis of
  variance problems, I. Effect of inequality of variance in the one-way
  classification}.
\newblock {\em The Annals of Mathematical Statistics}, 25(2):290--302.

\bibitem[Brueckner et~al., 2018]{Brueckner2018}
Brueckner, M., Titman, A., Jaki, T., Rojek, A., and Horby, P. (2018).
\newblock {Performance of different clinical trial designs to evaluate
  treatments during an epidemic}.
\newblock {\em PLoS ONE}, 13(9):e0203387.

\bibitem[Buitrago-Garcia et~al., 2020]{Buitrago-Garcia2020}
Buitrago-Garcia, D., Egli-Gany, D., Counotte, M.~J., Hossmann, S., Imeri, H.,
  Ipekci, A.~M., Salanti, G., and Low, N. (2020).
\newblock {Occurrence and transmission potential of asymptomatic and
  presymptomatic SARSCoV-2 infections: A living systematic review and
  meta-analysis}.
\newblock {\em PLoS Medicine}, 17(9):1--25.

\bibitem[Burnett et~al., 2020]{Burnett2020}
Burnett, T., Mozgunov, P., Pallmann, P., Villar, S.~S., Wheeler, G.~M., and
  Jaki, T. (2020).
\newblock {Adding flexibility to clinical trial designs: an example-based guide
  to the practical use of adaptive designs}.
\newblock pages 1--35.

\bibitem[Cake et~al., 2021]{Cake2021}
Cake, C., Ogburn, E., Pinches, H., Coleman, G., Seymour, D., Woodard, F.,
  Manohar, S., Monsur, M., Landray, M., Dalton, G., Morris, A.~D., Chinnery,
  P.~F., Hobbs, F.~R., and Butler, C. (2021).
\newblock Development and evaluation of rapid data-enabled access to routine
  clinical information to enhance early recruitment to the national clinical
  platform trial of covid-19 community treatments.
\newblock {\em medRxiv}.

\bibitem[Camacho et~al., 2015]{Camacho2015}
Camacho, A., Eggo, R.~M., Funk, S., Watson, C.~H., Kucharski, A.~J., and
  Edmunds, W.~J. (2015).
\newblock {Estimating the probability of demonstrating vaccine efficacy in the
  declining Ebola epidemic: A Bayesian modelling approach}.
\newblock {\em BMJ Open}, 5(12):1--6.

\bibitem[Chandereng and Chappell, 2019]{Chandereng2019}
Chandereng, T. and Chappell, R. (2019).
\newblock {Robust blocked response-adaptive randomization designs}.
\newblock {\em arXiv}.

\bibitem[Danon et~al., 2020]{Danon2020}
Danon, L., Brooks-Pollock, E., Bailey, M., and Keeling, M.~J. (2020).
\newblock {A spatial model of CoVID-19 transmission in England and Wales: early
  spread and peak timing}.
\newblock {\em medRxiv}, page 2020.02.12.20022566.

\bibitem[Dean et~al., 2019]{Dean2019}
Dean, N.~E., Gsell, P.-S., Brookmeyer, R., {De Gruttola}, V., Donnelly, C.~A.,
  Halloran, M.~E., Jasseh, M., Nason, M., Riveros, X., Watson, C.~H.,
  Henao-Restrepo, A.~M., and Longini, I.~M. (2019).
\newblock {Design of vaccine efficacy trials during public health emergencies}.
\newblock {\em Science Translational Medicine}, 11(499):eaat0360.

\bibitem[{Ebola {\c{c}}a Suffit Ring Vaccination Trial Consortium},
  2015]{Access2015}
{Ebola {\c{c}}a Suffit Ring Vaccination Trial Consortium} (2015).
\newblock {The ring vaccination trial: a novel cluster randomised controlled
  trial design to evaluate vaccine efficacy and effectiveness during outbreaks,
  with special reference to Ebola}.
\newblock {\em BMJ (Clinical research ed.)}, 351(July):h3740.

\bibitem[{European Centre for Disease Prevention and Control}, 2020]{ECDC20202}
{European Centre for Disease Prevention and Control} (2020).
\newblock {Resource estimation for contact tracing, quarantine and monitoring
  activities for COVID-19 cases in the EU/EEA}.
\newblock Technical Report March, European Centre For Disease Prevention And
  Control.

\bibitem[Friede and Kieser, 2002]{Friede2002}
Friede, T. and Kieser, M. (2002).
\newblock {On the inappropriateness of an EM algorithm based procedure for
  blinded sample size re-estimation}.
\newblock {\em Statistics in Medicine}, 21(2):165--176.

\bibitem[Fyles et~al., 2020]{Fyles}
Fyles, M., Fearon, E., and Working, M.~C. (2020).
\newblock {Household structured contact tracing: Branching process model}.
\newblock Technical report.

\bibitem[Gould and Shih, 1992]{Gould1992}
Gould, A.~L. and Shih, W.~J. (1992).
\newblock {Sample size re-estimation without unblinding for normally
  distributed outcomes with unknown variance}.
\newblock {\em Communications in Statistics - Theory and Methods},
  21(10):2833--2853.

\bibitem[He et~al., 2020]{He2020}
He, X., Lau, E.~H., Wu, P., Deng, X., Wang, J., Hao, X., Lau, Y.~C., Wong,
  J.~Y., Guan, Y., Tan, X., Mo, X., Chen, Y., Liao, B., Chen, W., Hu, F.,
  Zhang, Q., Zhong, M., Wu, Y., Zhao, L., Zhang, F., Cowling, B.~J., Li, F.,
  and Leung, G.~M. (2020).
\newblock {Temporal dynamics in viral shedding and transmissibility of
  COVID-19}.
\newblock {\em Nature Medicine}, 26(May).

\bibitem[Henao-Restrepo et~al., 2017]{Henao-Restrepo2017}
Henao-Restrepo, A.~M., Camacho, A., Longini, I.~M., Watson, C.~H., Edmunds,
  W.~J., Egger, M., Carroll, M.~W., Dean, N.~E., Diatta, I., Doumbia, M.,
  Draguez, B., Duraffour, S., Enwere, G., Grais, R., Gunther, S., Gsell, P.~S.,
  Hossmann, S., Watle, S.~V., Kond{\'{e}}, M.~K., K{\'{e}}{\"{i}}ta, S., Kone,
  S., Kuisma, E., Levine, M.~M., Mandal, S., Mauget, T., Norheim, G., Riveros,
  X., Soumah, A., Trelle, S., Vicari, A.~S., R{\o}ttingen, J.~A., and Kieny,
  M.~P. (2017).
\newblock {Efficacy and effectiveness of an rVSV-vectored vaccine in preventing
  Ebola virus disease: final results from the Guinea ring vaccination,
  open-label, cluster-randomised trial (Ebola {\c{C}}a Suffit!)}.
\newblock {\em The Lancet}, 389(10068):505--518.

\bibitem[Henao-Restrepo et~al., 2015]{Henao-Restrepo2015}
Henao-Restrepo, A.~M., Longini, I.~M., Egger, M., Dean, N.~E., Edmunds, W.~J.,
  Camacho, A., Carroll, M.~W., Doumbia, M., Draguez, B., Duraffour, S., Enwere,
  G., Grais, R., Gunther, S., Hossmann, S., Kond{\'{e}}, M.~K., Kone, S.,
  Kuisma, E., Levine, M.~M., Mandal, S., Norheim, G., Riveros, X., Soumah, A.,
  Trelle, S., Vicari, A.~S., Watson, C.~H., K{\'{e}}{\"{i}}ta, S., Kieny,
  M.~P., and R{\o}ttingen, J.~A. (2015).
\newblock {Efficacy and effectiveness of an rVSV-vectored vaccine expressing
  Ebola surface glycoprotein: interim results from the Guinea ring vaccination
  cluster-randomised trial}.
\newblock {\em The Lancet}, 386(9996):857--866.

\bibitem[Hitchings et~al., 2018]{Hitchings2018}
Hitchings, M.~D., Lipsitch, M., Wang, R., and Bellan, S.~E. (2018).
\newblock {Competing effects of indirect protection and clustering on the power
  of cluster-randomized controlled vaccine trials}.
\newblock {\em American Journal of Epidemiology}, 187(8):1763--1771.

\bibitem[Hitchings et~al., 2017]{Hitchings2017}
Hitchings, M. D.~T., Grais, R.~F., and Lipsitch, M. (2017).
\newblock {Using simulation to aid trial design: Ring-vaccination trials}.
\newblock {\em PLoS Neglected Tropical Diseases}, 11(3):1--12.

\bibitem[Hu and Rosenberger, 2006]{Hu20062}
Hu, F. and Rosenberger, W.~F. (2006).
\newblock {\em {The Theory of Response-Adaptive Randomization in Clinical
  Trials}}.
\newblock John Wiley \& Sons, Ltd., New Jersey.

\bibitem[Huang et~al., 2018]{Huang2018}
Huang, L., Bai, J., Yu, H., and Chen, F. (2018).
\newblock {Sample size re-estimation without un-blinding for time-to-event
  outcomes in oncology clinical trials}.
\newblock {\em Journal of Biomedical Research}, 32(1):23--29.

\bibitem[Hudgens et~al., 2004]{Hudgens2004}
Hudgens, M.~G., Gilbert, P.~B., and Self, S.~G. (2004).
\newblock {Endpoints in vaccine trials}.
\newblock {\em Statistical Methods in Medical Research}, 13(2):89--114.

\bibitem[Kahn et~al., 2018]{Kahn2018a}
Kahn, R., Rid, A., Smith, P.~G., Eyal, N., and Lipsitch, M. (2018).
\newblock {Choices in vaccine trial design in epidemics of emerging
  infections}.
\newblock {\em PLoS Medicine}, 15(8):1--12.

\bibitem[Kahn et~al., 2020]{Kahn2020}
Kahn, R., Villar, S.~S., and Lipsitch, M. (2020).
\newblock {Innovative Vaccine Trial Designs for EID Outbreak Response}.
\newblock In preparation.

\bibitem[Kennedy et~al., 2016]{Kennedy2016}
Kennedy, S.~B., Neaton, J.~D., Lane, H.~C., Kieh, M.~W., Massaquoi, M.~B.,
  Touchette, N.~A., Nason, M.~C., Follmann, D.~A., Boley, F.~K., Johnson,
  M.~P., Larson, G., Kateh, F.~N., and Nyenswah, T.~G. (2016).
\newblock {Implementation of an Ebola virus disease vaccine clinical trial
  during the Ebola epidemic in Liberia: Design, procedures, and challenges}.
\newblock {\em Clinical Trials}, 13(1):49--56.

\bibitem[Kiss et~al., 2017]{Kiss2017}
Kiss, I.~Z., Miller, J., and Simon, P. (2017).
\newblock {\em {Mathematics of Epidemics on Networks From Exact to Approximate
  Models}}.
\newblock Springer.

\bibitem[Kucharski et~al., 2020]{Kucharski2020}
Kucharski, A.~J., Klepac, P., Conlan, A., Kissler, S.~M., Tang, M., Fry, H.,
  Gog, J., and Edmunds, J. (2020).
\newblock {Effectiveness of isolation, testing, contact tracing and physical
  distancing on reducing transmission of SARS-CoV-2 in different settings}.
\newblock {\em medRxiv}, page 2020.04.23.20077024.

\bibitem[Li et~al., 2020]{Li2020}
Li, Q., Guan, X., Wu, P., Wang, X., Zhou, L., Tong, Y., Ren, R., Leung, K.~S.,
  Lau, E.~H., Wong, J.~Y., Xing, X., Xiang, N., Wu, Y., Li, C., Chen, Q., Li,
  D., Liu, T., Zhao, J., Liu, M., Tu, W., Chen, C., Jin, L., Yang, R., Wang,
  Q., Zhou, S., Wang, R., Liu, H., Luo, Y., Liu, Y., Shao, G., Li, H., Tao, Z.,
  Yang, Y., Deng, Z., Liu, B., Ma, Z., Zhang, Y., Shi, G., Lam, T.~T., Wu,
  J.~T., Gao, G.~F., Cowling, B.~J., Yang, B., Leung, G.~M., and Feng, Z.
  (2020).
\newblock {Early transmission dynamics in Wuhan, China, of novel
  coronavirus–infected pneumonia}.
\newblock {\em New England Journal of Medicine}, pages 1--9.

\bibitem[MacIntyre et~al., 2020]{MacIntyre2020.12.15.20248278}
MacIntyre, C.~R., Costantino, V., and Trent, M. (2020).
\newblock Modelling of covid-19 vaccination strategies and herd immunity, in
  scenarios of limited and full vaccine supply in nsw, australia.
\newblock {\em medRxiv}.

\bibitem[Marzi et~al., 2015]{Marzi2015}
Marzi, A., Robertson, S.~J., Haddock, E., Feldmann, F., Hanley, P.~W., Scott,
  D.~P., Strong, J.~E., Kobinger, G., Best, S.~M., and Feldmann, H. (2015).
\newblock {VSV-EBOV rapidly protects macaques against infection with the
  2014/15 Ebola virus outbreak strain}.
\newblock {\em Science}, 349(6249):739--742.

\bibitem[Nason, 2016]{Nason2016}
Nason, M. (2016).
\newblock {Statistics and logistics: Design of Ebola vaccine trials in West
  Africa}.
\newblock {\em Clinical Trials}, 13(1):87--91.

\bibitem[{Office for National Statistics}, 2011]{Statistics2011}
{Office for National Statistics} (2011).
\newblock {CT0819\_2011 Census - household type, household size and age of
  usual residents (households) - England and Wales}.

\bibitem[Pallmann et~al., 2018]{Pallmann2018}
Pallmann, P., Bedding, A.~W., Choodari-Oskooei, B., Dimairo, M., Flight, L.,
  Hampson, L.~V., Holmes, J., Mander, A.~P., Odondi, L., Sydes, M.~R., Villar,
  S.~S., Wason, J.~M., Weir, C.~J., Wheeler, G.~M., Yap, C., and Jaki, T.
  (2018).
\newblock {Adaptive designs in clinical trials: Why use them, and how to run
  and report them}.
\newblock {\em BMC Medicine}, 16(1):1--15.

\bibitem[Poland et~al., 2020]{Poland2020}
Poland, G.~A., Ovsyannikova, I.~G., and Kennedy, R.~B. (2020).
\newblock Sars-cov-2 immunity: review and applications to phase 3 vaccine
  candidates.
\newblock {\em The Lancet}.

\bibitem[Proschan and Evans, 2020]{Proschan2020}
Proschan, M. and Evans, S. (2020).
\newblock {Resist the temptation of response-adaptive randomization}.
\newblock {\em Clinical Infectious Diseases}, pages 1--8.

\bibitem[Rid and Miller, 2016]{RidMiller2016}
Rid, A. and Miller, F.~G. (2016).
\newblock Ethical rationale for the ebola ``ring vaccination'' trial design.
\newblock {\em Am J Public Health}, 106(3):432--5.

\bibitem[Rosenberger et~al., 2001]{Rosenberger2001}
Rosenberger, W.~F., Stallard, N., Ivanova, A., Harper, C.~N., and Ricks, M.~L.
  (2001).
\newblock {Optimal Adaptive Designs for Binary Response Trials With Three
  Treatments}.
\newblock {\em Biometrics}, 57:909--913.

\bibitem[Schoenfeld, 1983]{Schoenfeld1983}
Schoenfeld, D. A.~. (1983).
\newblock {Sample-size formula for the proportional-hazards regression model}.
\newblock {\em International Biometric Society}, 39(2):499--503.

\bibitem[Scott, 2020]{Scott2020}
Scott, I.~A. (2020).
\newblock {COVID-19 pandemic and the tension between the need to act and the
  need to know}.
\newblock {\em Internal Medicine Journal}, 50(8):904--909.

\bibitem[Shim and Galvani, 2012]{Shim2012}
Shim, E. and Galvani, A.~P. (2012).
\newblock {Distinguishing vaccine efficacy and effectiveness}.
\newblock {\em Vaccine}, 30(47):6700--6705.

\bibitem[Simon and Simon, 2011]{Simon2011}
Simon, R. and Simon, N.~R. (2011).
\newblock {Using randomization tests to preserve type I error with
  response-adaptive and covariate-adaptive randomization}.
\newblock {\em Statistics Probability Letters}, 81(7):767--772.

\bibitem[Stallard et~al., 2020]{Stallard}
Stallard, N., Hampson, L., Benda, N., Brannath, W., Burnett, T., Friede, T.,
  Kimani, P., Koenig, F., Krisam, J., Mozgunov, P., Posch, M., Wason, J.,
  Wassmer, G., Whitehead, J., Williamson, S., Zohar, S., and Jaki, T. (2020).
\newblock Efficient adaptive designs for clinical trials of interventions for
  covid-19.
\newblock {\em Statistics in Biopharmaceutical Research}, 12(4):483--497.

\bibitem[Tapiwa et~al., 2020]{Tapiwa2020}
Tapiwa, G., C{\'{e}}cile, K., Dongxuan, C., Andrea, T., Christel, F., Jacco,
  W., and Niel, H. (2020).
\newblock {Estimating the generation interval for COVID-19 based on symptom
  onset data}.
\newblock {\em medRxiv}, pages 1--13.

\bibitem[Teel et~al., 2015]{Teel2015}
Teel, C., Park, T., and Sampson, A.~R. (2015).
\newblock {EM estimation for finite mixture models with known mixture component
  size}.
\newblock {\em Commun Stat Simul Comput}, 44(6):1545--1556.

\bibitem[Thall and Wathen, 2007]{Thall2007}
Thall, P.~F. and Wathen, J.~K. (2007).
\newblock {Practical Bayesian adaptive randomization in clinical trials}.
\newblock {\em Eur J Cancer}, 43(5):859--866.

\bibitem[Tymofyeyev et~al., 2007]{Tymofyeyev2007}
Tymofyeyev, Y., Rosenberger, W.~F., and Hu, F. (2007).
\newblock {Implementing optimal allocation in sequential binary response
  experiments}.
\newblock {\em Journal of the American Statistical Association},
  102(477):224--234.

\bibitem[Villar et~al., 2015]{Villar2015a}
Villar, S.~S., Bowden, J., and Wason, J. (2015).
\newblock {Multi-armed bandit models for the optimal design of clinical trials:
  Benefits and challenges}.
\newblock {\em Statistical Science}, 30(2):199--215.

\bibitem[Villar et~al., 2020]{Villar2020}
Villar, S.~S., Robertson, D.~S., and Rosenberger, W.~F. (2020).
\newblock {The temptation of overgeneralizing response-adaptive randomization}.
\newblock {\em Infectious Diseases Society of America}.

\bibitem[Weinberg and Szilagyi, 2010]{Weinberg2010}
Weinberg, G.~A. and Szilagyi, P.~G. (2010).
\newblock {Vaccine epidemiology: Efficacy, effectiveness, and the translational
  research roadmap}.
\newblock {\em Journal of Infectious Diseases}, 201(11):1607--1610.

\bibitem[{WHO R\&D Blueprint}, 2020]{WHOR&DBlueprint2020}
{WHO R\&D Blueprint} (2020).
\newblock {An international randomised trial of candidate vaccines against
  COVID-19}.
\newblock Technical report.

\bibitem[Widdowson et~al., 2016]{Widdowson2016}
Widdowson, M.-A., Schrag, S.~J., Carter, R.~J., Carr, W., Legardy-Williams, J.,
  Gibson, L., Lisk, D.~R., Jalloh, M.~I., Bash-Taqi, D.~A., Kargbo, S. A.~S.,
  Idriss, A., Deen, G.~F., Russell, J.~B., McDonald, W., Albert, A.~P., Basket,
  M., Callis, A., Carter, V.~M., Ogunsanya, K. R.~C., Gee, J., Pinner, R.,
  Mahon, B.~E., Goldstein, S.~T., Seward, J.~F., Samai, M., and Schuchat, A.
  (2016).
\newblock {Implementing an Ebola Vaccine Study — Sierra Leone}.
\newblock {\em MMWR Supplements}, 65(3):98--106.

\bibitem[Williamson et~al., 2017]{Williamson2017}
Williamson, S.~F., Jacko, P., Villar, S.~S., and Jaki, T. (2017).
\newblock {Europe PMC Funders Group Europe PMC Funders Author Manuscripts A
  Bayesian adaptive design for clinical trials in rare diseases}.
\newblock pages 136--153.

\end{thebibliography}

\clearpage
\begin{appendix}

\section{The COVID-19 model}\label{modelC19}

Both the network model and the disease transmission model will be specific to a particular setting and, indeed, might change or become better informed over the course of the trial. Therefore, we use models without supposing that they will reflect the ``truth'' for any particular disease or scenario. Instead, we make choices we believe to be plausible in order to demonstrate the general methods, whose features will persist for settings such as these within the parameters we have described, i.e., diseases that pass from person to person, through contacts that can be traced.

\subsection{Network model}

\subsubsection{Definitions}

Our network is an undirected graph,
$$\mathcal{G}=(\mathcal{V,E})$$ with vertices, or nodes, $$\mathcal{V}=\{i,i=1,...,N_I\}$$ 
representing the $N_I$ individuals, which we index with $i$, and edges 
$$\mathcal{E}\subseteq\{\{i,j\}|(i,j)\in\mathcal{V}^2\cap i\neq j\}$$ 
representing connections, or relationships, between individuals. 
Each node $i$ has a set of attributes $\{a_1(i),a_2(i),...,a_{N_a}(i)\}$. 

The network operates at the level of the individuals. Each individual belongs to a household $\mathcal{H}_h$ and every individual in a household is connected to every other individual in the household, such that the induced subgraph $\mathcal{G}_1[\mathcal{H}_h]$ is completely connected: 
$$\mathcal{E}_1[\mathcal{H}_h]=\{\{i,j\}|(i,j)\in\mathcal{V}[\mathcal{H}_h]^2\cap i\neq j\}.$$
Let $\mathcal{H}=\{\mathcal{H}_h,h=1,...,N_H\}$ be the set of $N_H$ households, which forms a partition of the vertices $\mathcal{V}$ (all subsets are mutually disjoint and their union is equal to the set). Let $a_1(i)$ be the household of individual $i$, i.e. $a_1(i)=h\Leftrightarrow i\in\mathcal{V}[\mathcal{H}_h]$. Then the size of a household is
$$\mathcal{O}(\mathcal{H}_h) = \sum_{i=1}^{N_I} \textbf{1}_{a_1(i)=h}.$$

We start with  $N_H=500$ households with $\mathcal{O}(\mathcal{H}_h)$ people in each. The age and number distribution follows \citep{Statistics2011}. That is, each person has an age attribute, 
$$a_2(i)\in\{1,2,3\}$$
where $a_2(i)=1$ if person $i$ is under 19, $a_2(i)=2$ if person $i$ is 19 to 65, and $a_2(i)=3$ if person $i$ is over 65.

We define an individual $i$ to be part of the workforce via a ``worker'' variable $a_3(i)$, where 
\begin{align*}
a_2(i)=1 &\implies a_3(i)=0,\\
a_2(i)=2 &\implies a_3(i)=1,\\
a_2(i)=3 &\implies \left\{\begin{array}{lr}a_3(i)=0 & \text{ with probability } 0.8\\
a_3(i)=1 & \text{ with probability } 0.2.\end{array}\right.
\end{align*}
There are $A_3=\sum_i\textbf{1}_{a_3(i)=1} \approx 731$ people in the workforce. We define $A_3/15\approx 49$ workplaces to which people with $a_3(i)=1$ (that is, all those aged 19 to 65, and 1/5 of people over 65) are assigned following a multinomial distribution. We choose to place on average 15 people in a ``workplace'', $\mathcal{W}_w$, which represents not their employment structure but a close shared use of the infrastructure. A useful guide might be how many toilets per person a place of work should have.

The workplaces are completely connected: 
$$\mathcal{E}[\mathcal{W}_w]=\{\{i,j\}|(i,j)\in\mathcal{V}[\mathcal{W}_w]^2\cap i\neq j\}.$$

Finally, 1000 totally random edges are added, which amounts to around ten per person:
$$\text{Pr}\left(\{i,j\}\in\mathcal{E}\right)=\frac{10}{\left(\sum_h\mathcal{O}(\mathcal{H}_h)\right)}.$$
These edges correspond to potential transmission encounters that would not be recalled or anticipated through contact tracing. The result is an average of 20 connections per person, of which ten have a weight of 1 and ten have a weight of 0.1.

\clearpage

\subsubsection{Parametrisation}

The parameters we have used and their provenance are listed in Table \ref{parameters}.

\begin{table}[h!]
\centering
\caption{Parameters used in the COVID-19 disease transmission and vaccine trial model. $\mathcal{N}(l,\mu,\sigma)$ denotes a normal distribution truncated at $l$.}
\begin{tabular}{|m{5cm}|m{5cm}|m{5cm}|}
\hline
Parameter  & Value & Source  \\
\hline
Number of households & 500 & Chosen with reference to other choices \\
Household size & Draws from raw data & UK 2011 census \\
Workplace size & $\sim\text{Poisson}(15)$ & \url{https://www.hse.gov.uk/contact/faqs/toilets.htm} \\
\hline
$\beta$ (per-contact infection or transmission rate) & 0.01 & Chosen with reference to other choices \\ 
Known edge weight ($\chi_h$, $\chi_w$) & 1 & Chosen with reference to other choices \\
Unknown edge weight  ($\chi_n$) & 0.1 & Chosen with reference to other choices \\
\hline
{SARS-CoV-2} incubation period $\xi$ & $\sim 2+\Gamma(\text{shape}=13.3,\text{rate}=4.16)$ & \citet{Li2020} \\
{SARS-CoV-2} infectious period $\gamma$ & $\sim 1+\Gamma(\text{shape}=1.43,\text{rate}=0.549)$ & \citet{Li2020} \\
Time to enrol whole contact network & $\sim\mathcal{N}(l=0,10.32,4.79)$
&  \citet{Henao-Restrepo2017}\\
Time to vaccine-induced seroconversion & $\sim\Gamma(\text{shape}=3,\text{rate}=1)$ & Plausibly all seroconverted within 14 days \\
\hline
\end{tabular}
\label{parameters}
\end{table}

\clearpage

\subsection{Disease and trial state transitions}

\subsubsection{Disease transmission rules}

A susceptible individual $i$ belonging to arm $x$ becomes ``exposed'' (i.e. transitions to state $E$) with rate 
$$k_{x}(i)=\beta \left(\chi_h\mathcal{M}_{i}^{(h)}+\chi_h\mathcal{L}_{i}^{(h)}+\chi_w\mathcal{L}_{i}^{(w)}+\chi_n\mathcal{L}_{i}^{(n)}\right)(1-\eta\cdot x_i),\quad x\in\{V,U,C\},$$ 
where $h$ refers to household, $w$ to workplace, and $n$ to random.  $\chi_h>0$ is the edge weight for household contacts, $\chi_w>0$ is the  edge weight for workplace contacts, and  $\chi_n>0$ is the edge weight for random contacts. One could specify a different $\chi_h$ for pre- and post-symptomatic infectiousness, or even a changing profile over time, as described in \citet{He2020}.  $x_i=1$ if person $i$ is vaccinated and 0 otherwise. {That is, we assume that the vaccine prevents infection, and does not change infectiousness or the likelihood of asymptomatic disease among those who become infected after vaccination or a prophylactic effect among those who were infected before.} 
$\eta\leq 1$ is the vaccine efficacy, and $\beta>0$ is the per-contact rate at which infectious people infect their susceptible contacts.\footnote{The unit of $\beta$ is per contact per day; the edge weights $\chi$ are unitless. They can be thought to transform the static (or quenched) network into a dynamic or adaptive network through link deactivation \citep{Kiss2017}.} $\mathcal{M}_{i}^{(h)}$ is the number of $I_S$ household ($h$) contacts of susceptible $i$.
$$\mathcal{L}_{i}^{(y)}=\sum_{j\in\mathcal{N}_{(y)}(i)}\textbf{1}_{j\in\{I_{A,\cdot},I_{P,\cdot}\}},\quad y\in\{h,w,n\}$$
is similarly defined as the number of contacts who are infectious but not symptomatic.

Note that there is no infection other than from a contact (with the exception of the simulated ``time trends'', Figure \ref{time_trend_fig}). Thus any infection within a contact network came (directly or indirectly) from the contact network's index case.

Our model and parameter choices yield new infection events as occurring from pre-symptomatic people 58\% of the time and from symptomatic people 42\% of the time, omitting transmissions from people who never become symptomatic. In comparison to reported estimates (44\% from 77 recorded transmission pairs \citep{He2020}, and 48 and 62\% in Singapore and Tianjin, respectively \citep{Tapiwa2020}, we confirm that our simulation scenario is consistent with one with quick quarantine of close contacts \citep{He2020}.

\subsubsection{Trial rules}

In our trial we define contact tracing as identifying only existing relationships (``acquaintances'', or ``those met before'', in the terminology of \citet{Kucharski2020}) that might be, or might have been, a means for transmission. That is, a newly diagnosed person is asked to recall all of their contacts, and these individuals are contacted and asked likewise to list all their contacts. This makes our simulation and trial design most like the ``self-isolation and manual contact tracing of acquaintances'' of \citet{Kucharski2020}.  The relationships that are of interest will depend on the society and any concurrent actions, guidance or instruction from the state, which in our simulation include home and workplace contacts, while random contacts remain unknown. We assume unknown relationships are not recalled in contact tracing and, similarly, cannot be anticipated for recruitment of trial participants. We assume work and home relationships are recalled perfectly. By recruiting ``contacts'' and ``contacts of contacts'' into the trial, we are recruiting the people the index case lives with and their colleagues, and those the case works with and their housemates. 

For our purposes, an index case for a contact network is a person identified as being in state $I_{S,\cdot}$ after the initiation of the trial. Eligible people are traced as described in \citet{ECDC20202} and those who consent are enrolled as soon as they are identified and give their consent. { It takes on average ten days from symptom onset to enrol a contact network, which includes the time to report, the time to contact trace, and the time to enrol \citep{Henao-Restrepo2017}.} Susceptible and exposed people are eligible for enrolment if they are a ``known'' contact of the index case and they are not already enrolled in the trial. Symptomatic $I_{S,U}$ people are excluded on the basis of their symptoms. $R_{U}$ are excluded on the basis of their history, which we assume a perfect knowledge of in our simulation. This could result either from people being able to identify having had COVID-19, or from there being an accurate and reliable antibody test. Inclusion of $R_U$ people in the simulated trial would result in a dilution of infections and would therefore require enrolment of more participants to maintain power. Their inclusion is advocated for reasons of safety testing (\url{https://www.who.int/publications/i/item/an-international-randomised-trial-of-candidate-vaccines-against-covid-19}, \url{https://clinicaltrials.gov/ct2/show/NCT04405076}). Elsewhere seropositive people are excluded from vaccine trials (\url{https://www.clinicaltrialsregister.eu/ctr-search/trial/2020-001228-32/GB}).

 The enrolment rate is $0\leq \epsilon\leq 1$, which is the probability for each eligible person to enrol, where we use only contact structure and lack of symptoms to define eligibility. We use $\epsilon=0.7$.\footnote{\citet{Henao-Restrepo2017} report that 50\% of people identified were eligible and enrolled, where their eligibility criteria excluded people who were pregnant, breastfeeding, or under the age of 18.} Enrolled participants are randomised to the experimental arm with probability $0\leq \pi_1\leq 1$, and  $\pi_1+\pi_0=1$. Enrolment { following symptom onset for the index case} takes time $\alpha$, {which we describe following observed enrollment times  \citep{Henao-Restrepo2017}. Alternatives include a Poisson distribution with parameter $\sim$ Uniform(1.5, 2.5) \citep{Fyles}. Ideally, this parameter would be determined according to observed contact-tracing efforts and the expected capacity of the trial team.} 
 
 Those vaccinated have an additional wait time before reaching state $S_V$, which is development of immunity (or time to seroconversion or detectable antibodies), and which takes time $\tau$. Transition from $S_U$ to $E_U$ is possible in this wait time.  We assume that the vaccine effect before seroconversion is zero and that the vaccine effect after is the full effect of the vaccine.

Individuals who are in state $E$ and are unenrolled are enrolled with the same probability ($\epsilon$) as the susceptibles $S_U$, as they are asymptomatic. The same wait time for enrolment applies, but time to seroconversion does not, as the individual is already infected. If the participant transitions to $I_{S,U}$ before the recruitment time elapses, they will be excluded from the trial, as they will be showing symptoms and can be tested for confirmation. 

Result accrual relies on surveillance and self reporting. For our simulations we assume that a fraction $1-\delta=0.2$ of infectious individuals are asymptomatic. These infection events go unreported in the trial results (but still contribute to onward transmission). {Our simulation mirrors the rules given to the participants: that upon the onset of symptoms, they should stay at home, and notify the trial team, who will organise the PCR test. In this way, the day of symptom onset is recorded (to be confirmed by PCR). We assume that this information will be available at the end of the follow-up period (\fuday days). This means that we assume that a person who becomes symptomatic on day \fuday is tested and confirmed on the same day. In our simulation we assume that the rules are followed without error.}

\bigskip

We simulate one network at a time, beginning when the index case is identified. Transmission in each network is independent of all other networks in the trial. The networks are related only through the time reference, in that one contact network is initiated on each day. 

Finally, we assume that randomisation, enrolment and vaccination all happen on a single day for each individual, although the day will differ between individuals.

\clearpage

\section{Analyses, and exclusion criterion implemented at analysis points}\label{exclusion_app}

Here we detail all equations to accompany Section \ref{weighting_app}, which describes the different ways we explore to analyse the outcome. We present the methods in the same order and use a single framework that describes all the methods in the same way.

\subsection{Analysis of raw data}

We have $j=1,...,N_I$ individuals. Each has a vaccination status, $x_j$, and a disease status $y_j$, where the vaccination status is dictated by the trial design and the disease status from the underlying epidemic model:
$$x_j=\left\{\begin{array}{lr} 0 & \text{person $j$ not vaccinated}\\ 1 & \text{person $j$ vaccinated} \end{array} \right.$$
and
$$y_j=\left\{\begin{array}{lr} 0 & \text{person $j$ not diagnosed}\\ 1 & \text{person $j$ diagnosed} \end{array} \right.$$
at the end of the trial. 

The test statistic is a standard normal variable $Z$, where
\begin{equation}\label{zeq}Z=\frac{\hat{p}_1-\hat{p}_0}{\sqrt{\sigma_0+\sigma_1}},\end{equation}
\begin{equation}\label{peq}\hat{p}_v=\frac{\sum_{j:x_j=v,y_j=0}\omega_j}{N_v},\end{equation}
\begin{equation}\label{neq}N_v=\sum_{j:x_j=v}\omega_j\end{equation}
$$\omega_j=1 \quad \forall j, $$
and
\begin{equation}\label{seq}\sigma_v=\frac{\hat{p}_v(1-\hat{p}_v)}{N_v}.\end{equation}
$p_v$ is the true probability of not being a confirmed case if in arm $v$, and $\hat{p}_v$ is our estimate of it, defined as the proportion of people in arm $v$ not confirmed, where $v=0$ is the control arm and $v=1$ is the experimental arm; $N_v$ is the total number in arm $v$; and $\sigma_v$ is the variance of the estimator $\hat{p}_v$. Power is defined as the proportion of $Z$ values that exceed 1.64, which is the 95th quantile of a standard normal distribution. $\omega_j$ is the weight of person $j$ which, for the unweighted method, is 1 for all participants. In the descriptions that follow, we see that the weights $\omega$ define the retrospective exclusion criterion so that all methods use the same calculation and each is defined only by the definition of the weights. 

\subsection{Analysis using binary weighting}

The binary-weighting method proceeds as above but considers also the day of commencement of symptoms, $s_j$. Individuals are excluded if $s_j<9$ relative to a randomisation day of 0. We write this as the weight, $\omega_j$, for each individual $j$, so that a weight of 0 equates to exclusion:
$$\omega_j=\left\{\begin{array}{lr} 0 & s_j<9\\ 1 & s_j\geq 9\text{ or }y_j=0\end{array} 
\right.$$

These are used together with Equations  \ref{zeq}, \ref{peq}, \ref{neq}, and \ref{seq} as before.

\subsection{Analysis using continuous weighting}

Given person $j$'s symptoms began on day $s_j$ relative to their randomisation day of 0, $\tau$ and $\xi$ are their unknown time to seroconversion\footnote{Note that we include time to seroconversion also for the control group, who don't receive the vaccine, and don't seroconvert.} and incubation time, respectively. The probability they were infected after the trial began is $P(\tau+\xi<s_j)$. We assume $\tau$ and $\xi$ are distributed between individuals as $\Gamma(\text{shape}=3,\text{rate}=1)$ and $2+\Gamma(\text{shape}=13.3,\text{rate}=4.16)$. Hence the distribution of $\tau+\xi$ is estimated by matching moments using the Welch--Satterthwaite equation, as described in \citet{Box1954}.

\begin{figure}[ht]
\centering
\includegraphics[width=0.6\textwidth]{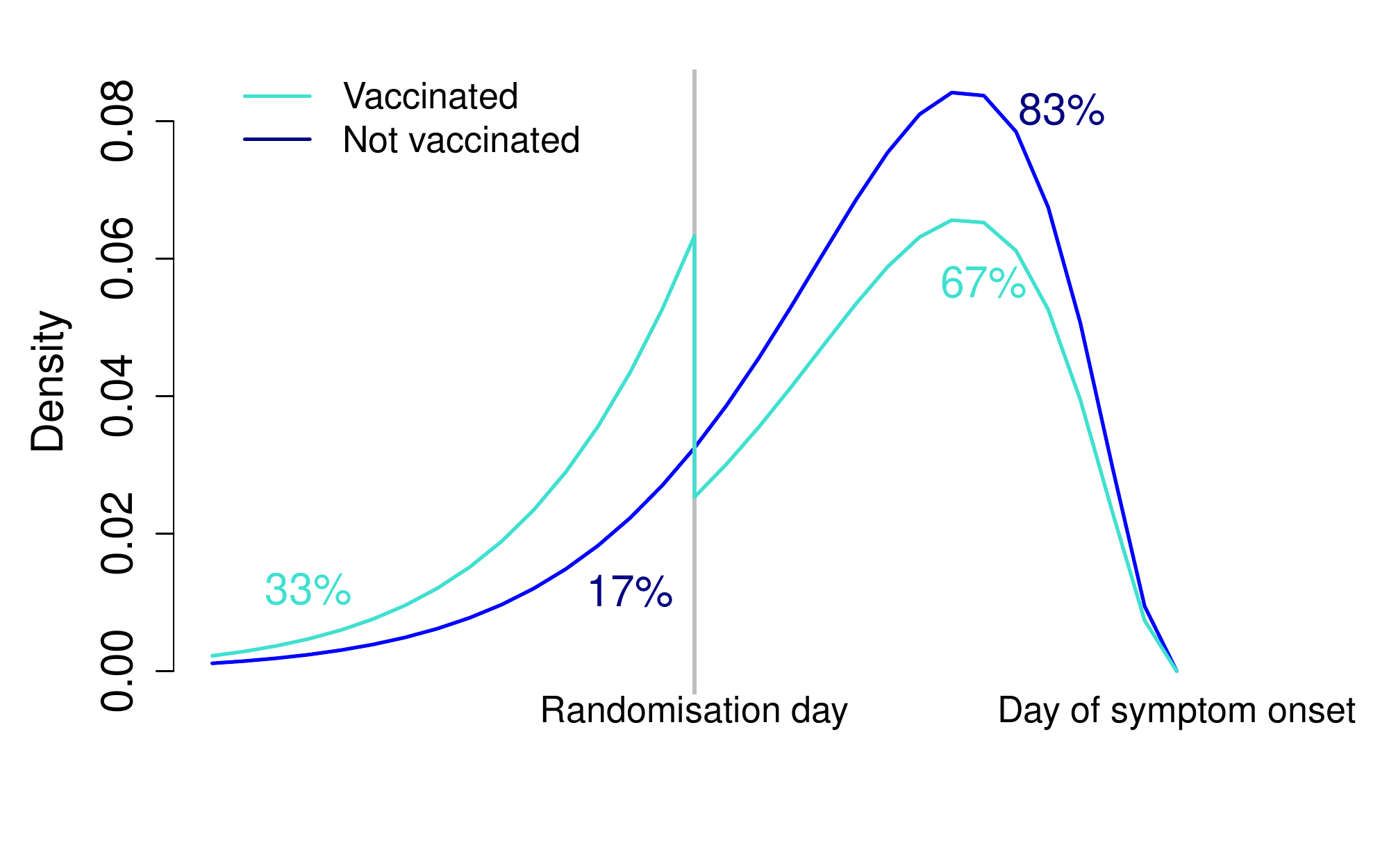}
\caption{\small The curves show the probability that the value on the $x$ axis was the day a person became infected, given the day they showed symptoms, whether or not they were vaccinated, and the effect of the vaccine (where we omit vaccine-induced antibody response for simplicity of presentation). The inclusion weight given the day of symptom onset is the area under curve to the left of the randomisation day. Two participants show symptoms on the same day. One (turquoise) is vaccinated. The other (navy) is not. Given the distribution of the incubation period, the navy person has a probability of 0.83 of having been infected after randomisation day. Their weight is therefore 0.83. Given that the vaccine efficacy estimate exceeds zero, the turquoise person is less than 83\% likely to have been infected after randomisation day -- 67\%, for an estimated vaccine efficacy of 0.6. In the binary-weighting case, both these participants have weight 0 (if day of symptom onset is less than the threshold of nine days) or weight 1 (otherwise).
}
\label{veweight}
\end{figure}

We estimate the vaccine efficacy $0\leq \hat{\eta} \leq 1$ as
$$\hat{\eta} = 1 - \left.\cfrac{f_1}{N_1}\right/\cfrac{f_0}{N_0},$$
where
$$\hat{f}_v=\sum_{j:x_j=v,y_j=1}\omega_j.$$
Suppose from the cumulative distribution function of the gamma distribution we have a nominal probability, i.e. neglecting the effect of the vaccine, $q_j$ that the day person $j$ was infected, $d_j$, was after seroconversion on day $D_c$. We write the complement, the probability that person $j$ was infected before day $D_c$, as $r_j=1-q_j$. We re-estimate their probability given that person $j$ was vaccinated ($x_j=1$):
\begin{equation}
\text{Pr}(d_j>D_c|x_j=1) = \frac{(1-\hat{\eta}) x_{j}}{y_{j}+(1-\hat{\eta}) x_{j}},
\end{equation}
because, if there is some efficacy, then they are more likely to have been infected before being vaccinated than after (relative to a vaccine that has no effect: $\text{Pr}(d_j>D_c|x_j=0) = q_j$). We solve this iteratively for $\hat{\eta}$ with reference to all observations $j$.\footnote{This is solved as in expectation maximisation (EM). EM algorithms have been used and discussed in clinical trials for e.g. sample-size re-estimation \citep{Gould1992,Friede2002,Teel2015,Huang2018}.} Then 
\begin{equation}
\omega_j=\left\{\begin{array}{lr}
1 & y_j=0 \\
\text{Pr}(d_j>D_c|x_j) & y_j=1
\end{array}\right.\end{equation}
See Figure \ref{veweight} for an illustration of the relationship between weight and vaccine effect. The resulting weights are used in Equations  \ref{zeq}, \ref{peq}, \ref{neq}, and \ref{seq} as before. 

\subsubsection*{Determining the inclusion weight for a vaccinated person}

The inclusion weight for a vaccinated person $j$ ($q_j=1$), whose symptoms began after $D_c$ (randomisation date plus an assumed time from vaccination to seroconversion) is the probability that they were infected after $D_c$.  To determine this, first we make explicit the condition that their infection date $d_j$ is less than their symptom date $D_s$.  The conditional probability we want can then be decomposed into probabilities unconditional on the symptom date, as follows,
\[
P(d_j > D_c | q_j = 1, d_j < D_s) = P(D_c < d_j < D_s, q_j = 1) / P(d_j < D_s, q_j = 1)
\]
Then splitting the denominator into the probabilities of being infected in two different periods (before versus after $D_c$) gives
\[
P(d_j > D_c | q_j = 1, d_j < D_s) = \frac{P(D_c < d_j < D_s | q_j = 1)}{P(d_j < D_c | q_j = 1) + P(D_c < d_j < D_s | q_j = 1)}
\]
Since the probability of being infected before $D_c$ doesn't depend on whether person $j$ was vaccinated, $P(d_j < D_c | q_j = 1) = P(d_j < D_c | q_j = 0)$.   We additionally assume that
\[
P(D_c < d_j < D_s | q_j = 1) =  \psi P(D_c < d_j < D_s | q_j = 0)
\] 
where $\psi$ is the relative risk of infection between a vaccinated and unvaccinated person, assumed to be constant through time, giving
\[
P(d_j > D_c | q_j = 1, d_j < D_s) = \frac{\psi  P(D_c < d_j < D_s | q_j = 0)}{P(d_j < D_c | q_j = 0) + \psi  P(D_c < d_j < D_s | q_j = 0)}
\]
Expressing the right hand side in terms of probabilities conditional on symptoms, by dividing the numerator and denominator by $P(d_j  < D_s | q_j  = 0)$, then gives the weight for a vaccinated person $j$ as
\[
\omega_j = P(d_j > D_c | q_j = 1, d_j < D_s) = \frac{\psi  \omega_j^{(0)}}{(1 - \omega_j^{(0)}) + \psi \omega_j^{(0)}}
\]
where $\omega_j^{(0)} = P(D_c < d_j < D_s | d_j < D_s, q_j = 0) =  P(D_c < d_j < D_s | q_j=0) / P(d_j  < D_s | q_j  = 0)$ is the weight if person $j$ were unvaccinated.

\clearpage

\section{Supplementary results}\label{suppres}

\subsection{Dependence of ring recruitment on contact tracing}\label{recruitment_appendix}

As ring recruitment relies on contact tracing, here we present a sensitivity analysis that demonstrates how the success of the method depends on the ``known fraction''. We define the ``known fraction'' of a network as $r_s = e_s w_s / (e_s w_s + e_u w_u)$, the weighted sum of known edges over the total weighted sum of edges, where there are $e_s$ known and $e_u$ unknown edges, with relationship weights $w_s$ and $w_u$ respectively. The net result is that the known fraction is the proportion of new cases that are in the contact networks of recent cases. Therefore the known fraction is entirely determined by how ``contact tracing'' is implemented: any contact that is not traced is by definition transient. The known fraction is 0.91 for our standard network (see Appendix \ref{modelC19}).

In Table \ref{ratio} we consider three possible values for the known fraction $r_s$ by holding the edges fixed and adjusting the relationship weights. To show power as a function of the known fraction, we fix the trial size to be 100 contact networks of participants. The results show the extent to which power diminishes as the fraction of known transmissions decreases.  As contacts become more ``known'' (thus more contacts are recruited into the trial), the number of confirmed cases within the trial increases, especially in the ``null case'' where there is no vaccine effect (VE=0). This is a result of the recruitment method better targeting those at risk.

Where it is possible to predict who is at imminent risk of infection, ring recruitment designs are advantageous compared to random recruitment. We expect to see a high known fraction $r_s$ for diseases whose transmission depends on exchange of or exposure to bodily fluids, such as EVD, as well as for COVID-19 in societies under ``lockdown'', where public spaces are closed and people stay at home, and/or there is extensive quarantining. We expect that the more restrictions to movement and activity there are, the greater the fraction of known transmission events. 

\newgeometry{,vmargin=2cm,hmargin=1cm}
\begin{landscape}
\begin{table}[ht]
\centering
\caption{\small The effect of the known fraction of transmission events on power. The trial follows the FR design with a follow-up time of \fuday days and ends once 100 contact networks have been enrolled. Participants are recruited following the ring strategy. Standard deviations  {for 10,000 simulations} in brackets.   }
\label{ratio}
\begin{tabular}{p{2cm}p{2cm}p{1.7cm}rrrp{2cm}p{2cm}}
  \hline
Known fraction $r_s$  & Number of participants & Number of confirmed cases & Vaccinated & Power & Type 1 error & Number of confirmed cases (VE=0) & VE estimate \\ 
  \hline
1.0 & 2910 (170) & 83 & 1454 & 0.91 & 0.04 & 106 & 0.59 (0.14) \\ 
  0.6 & 2907 (170) & 45 & 1452 & 0.60 & 0.05 & 55 & 0.55 (0.24) \\ 
  0.2 & 2907 (170) & 19 & 1452 & 0.26 & 0.05 & 22 & 0.43 (0.62) \\ 
   \hline
\end{tabular}
\end{table}
\end{landscape}
\restoregeometry

\subsection{Time trend in incidence of infection}

\begin{figure}[ht]
\centering
\includegraphics[width=0.95\textwidth]{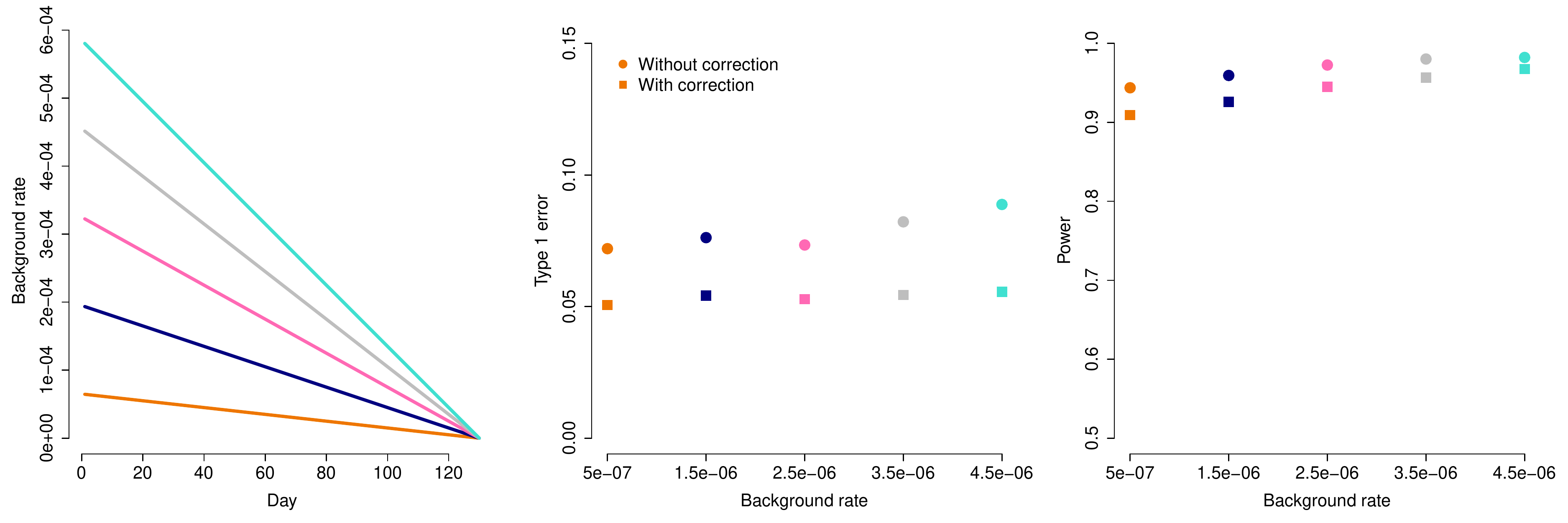}
\caption{The robustness of one adaptive trial design (TST) to a time trend, demonstrating the effect of the correction of \citet{Simon2011}. Type 1 error rate and power as a function of the trend in the background rate for a trial with a response-adaptive randomisation rate. ``Background rate'' can be interpreted as the rate of infection by individuals unknown in the context of the trial (e.g. source population or unknown contact). We illustrate with linear trends rather than something more realistic, undulating, or stochastic in order to stress test the method. In addition, we choose a trend that we would expect to most favour the Thompson sampling methods. As TS is a more ``aggressive'' response-adaptive method than TST, we expect it to be more susceptible to time trends, and therefore the loss in power following correction to be exaggerated. Left: Five different time trends for the background rate. The trend is that background rates diminish over time to zero. Middle: The gradient of the trend is shown on the x axis. On the y axis is the type 1 error rate. Right: The gradient of the trend is shown on the x axis. On the y axis is the power. }
\label{time_trend_fig}
\end{figure}

Figure \ref{time_trend_fig} shows the robustness of one adaptive trial design to a time trend, demonstrating the effect of the correction of \citet{Simon2011}. We choose to illustrate with linear trends rather than something more realistic, undulating, or stochastic in order to stress test the method. In addition, we choose a trend that we would expect to most favour the Thompson sampling methods. Indeed, we find that uncorrected Thompson sampling sees an increase in type 1 error, which is corrected by the resampling method of \citet{Simon2011}. As TS is a more ``aggressive'' response-adaptive method than TST, we expect it to be more susceptible to time trends, and therefore the loss in power following correction to be exaggerated. 

\subsection{Allocation probabilities}

\begin{figure}[ht]
\centering
\includegraphics[width=0.9\textwidth]{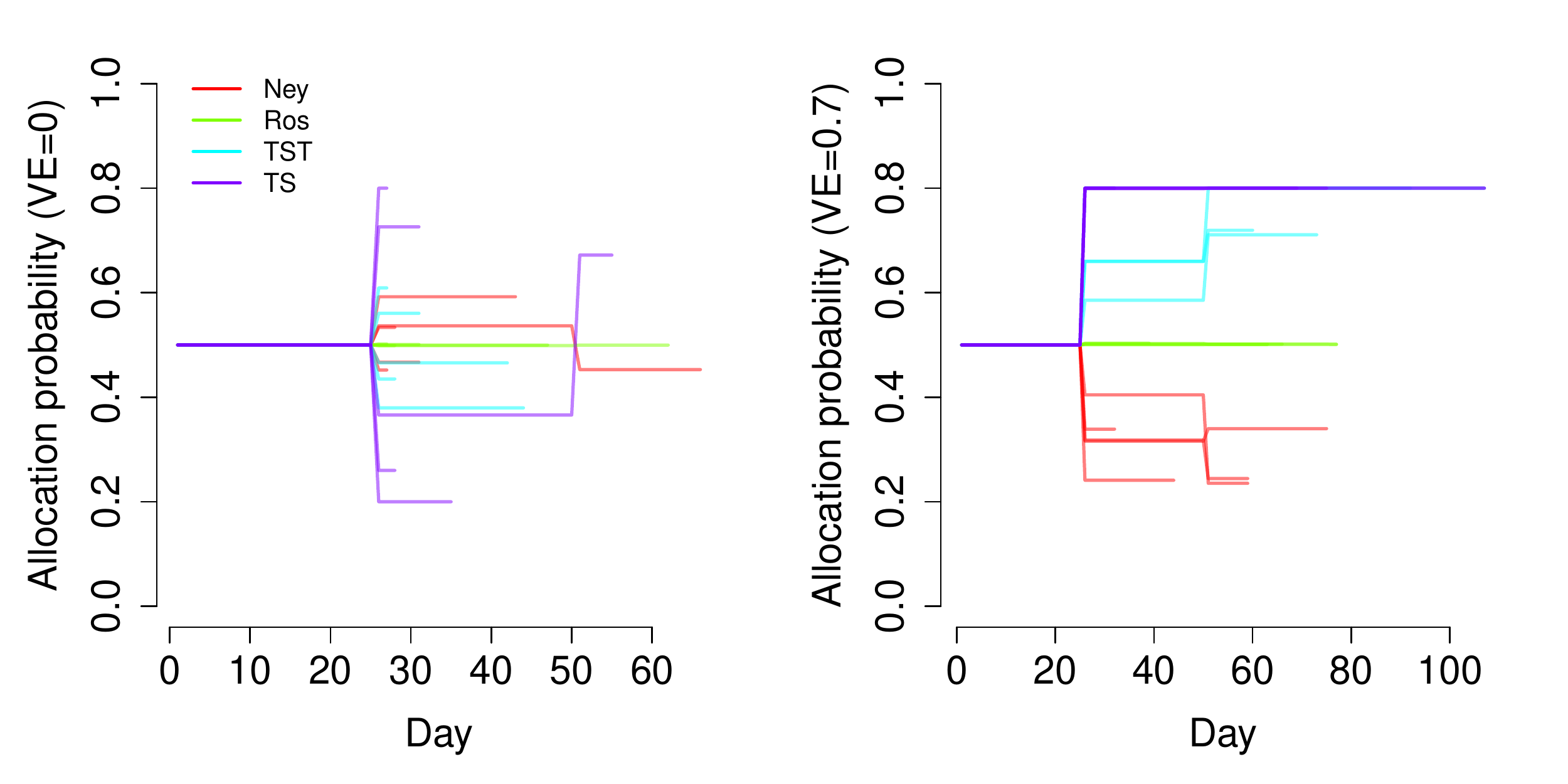}
\caption{Five samples of trajectories of the allocation probability over time for the adaptive designs, correponding to Table \ref{rar}. }
\label{allocation_probability}
\end{figure}

\subsection{Thompson sampling methods without early stopping}

In Table \ref{rarts}, we show the Thompson sampling methods again, where we do not conclude the trial early for efficacy. This results in much longer trials: on average 6 and 17 days longer (for TST and TS, respectively) than the trials that do stop early, and much harsher power corrections due to the extended periods spent at unequal randomisation, in comparison to Table \ref{rar}.

\subsection{Fixed-duration trials}\label{fixed_trial_duration_section}
We additionally compare the response-adaptive trials when there is a strict limit on their duration and there exists an efficacious vaccine, for example for a trial that lasts at most 85 days, and we evaluate the trial impact in terms of total people vaccinated up to 65 days after the trial's end. First we consider that the same number of people ($N_P = 32$) can be vaccinated per day, and then that ten times as many ($N_P = 320$) can be. Table \ref{60} compares the designs after { the total duration of} 150 days, where the trial has a fixed duration of at most 85 days{, and the number of people vaccinated in the remaining time depends on the probability that the trial successfully identified that the vaccine was efficacious, and the number of people that can be vaccinated per day post-trial}. With limited time, the trials that allocate most participants to the efficacious experimental arm -- TS and TST -- would be preferable in terms of vaccinating the most people during the trial, and up to day 150, if the rate of vaccination after the trial is the same as the rate of enrolment into the trial. These methods also see fewer exported infection events in the course of the trial -- both in total and per day. However, if, after the trial, ten times as many people can be vaccinated per day as were enrolled in the trial, the designs with higher powers are preferable, as they are better able to identify an efficacious vaccine, and therefore will vaccinate more people in the period after the trial.


\newgeometry{,vmargin=2cm,hmargin=1cm}

\begin{landscape}

\begin{table}[h]
\centering
\caption{Thompson-sampling response-adaptive designs where the trial design does not include early stopping. The outcome has a continuous  weighting. Participants are recruited following the ring strategy. The final follow-up time is \fuday days. The trial ends when \targetweight effective cases have been observed. Standard deviations in brackets. Correction for time trend uses the resampling method of \citet{Simon2011}.}
\label{rarts}
\begin{tabular}{lp{1.7cm}p{1.7cm}p{1.7cm}rrp{1.7cm}p{1.7cm}p{1.7cm}lp{1.7cm}}
  \hline
Adaptation  & Number of participants & Duration (days) & Number of confirmed cases & Vaccinated & Power & Power (corrected) & Type 1 error & Type 1 error (corrected) & VE estimate & Prevented exported infections \\ 
\hline
  TST & 2238 (712) & 94 (22) & 56 & 1423 & 0.76 & 0.70 & 0.04 & 0.04 & 0.62 (0.19) & 6.33 \\ 
  TS & 2346 (745) & 98 (23) & 57 & 1582 & 0.74 & 0.47 & 0.04 & 0.06 & 0.62 (0.19) & 6.60 \\ 
   \hline
\end{tabular}
\end{table}

\begin{table}[h]
\centering
\caption{\small Comparison of response-adaptive trials that last at most 85 days. We compare their profiles in terms of the totals vaccinated up to day 150{, which includes a maximum of 85 days of vaccinations as part of the trial and a minimum of 65 days of vaccination roll-out given a successful trial}. ``Vaccinated in trial'' is the expected number vaccinated during the trial, and ``Vaccinated up to day 150: assuming \ppday (320) per day'' is the expected number vaccinated after the trial, and up to day 150, estimated as \ppday (320) people vaccinated per day, for the days remaining after the end of the trial, multiplied by the power (the probability to have concluded efficacy and rolled out the vaccine --- bold text indicates whether uncorrected or drift-corrected power is used). Standard deviations  {for 10,000 simulations} in brackets. Correction for time trend uses the resampling method of \citet{Simon2011}.}
\label{60}
\begin{tabular}{lllp{2.4cm}p{1.5cm}rp{1.7cm}p{2.45cm}|p{1.6cm}p{1.8cm}}
  \hline
  &&&&&&&& \multicolumn{2}{c}{Vaccinated up to day 150:} \\
Adaptation & Sample size & Duration & Number of confirmed cases & Vaccinated in trial & Power & Power (corrected) & Exported infections in trial\footnotemark[2] & assuming 32 per day & assuming 320 per day \\ 
  \hline
  
  Ney. & 1937 (117) & 85 (4) & 54 & 808 & 0.78 & \textbf{0.73} & 24 & 2,345 & 16,065 \\ 
  Ros. et al. & 1938 (116) & 85 (4) & 51 & 976 & \textbf{0.77} & 0.74 & 24 & 2,524 & 16,905 \\ 
  TST & 1783 (268) & 80 (8) & 45 & 1098 & 0.73 & \textbf{0.71} & 21 & 2,696 & 16,953 \\ 
  TS & 1615 (433) & 75 (13) & 40 & 997 & 0.72 & \textbf{0.64} & 20 & 2,549 & 16,398 \\ 
  FR & 1938 (116) & 85 (4) & 52 & 969 & \textbf{0.77} &  & 24 & 2,582 & 16,973 \\ 
   \hline
\end{tabular}
\end{table}

\end{landscape}
\restoregeometry

\end{appendix}

\end{document}